\documentclass[a4paper,11pt]{article}
\usepackage{jheppub}
\usepackage[T1]{fontenc}
\ifPDFTeX\usepackage[utf8]{inputenc}\fi
\usepackage{lmodern,amsmath,amssymb,mathrsfs,xcolor,microtype}
\numberwithin{equation}{section}
\allowdisplaybreaks[2]
\renewcommand{\afterTocRuleSpace}{\clearpage}
\renewcommand{\d}[1]{\ensuremath{\operatorname{d}\!{#1}}}
\definecolor{palatinateblue}{rgb}{0.15,0.23,0.89}
\definecolor{brightpink}{rgb}{1.0,0.0,0.5}
\definecolor{amaranth}{rgb}{0.9,0.17,0.31}
\hypersetup{linktoc=all,colorlinks,linkcolor=palatinateblue,citecolor=brightpink,urlcolor=amaranth}

\begin{document}

\title{Charged AdS$_{3}$ black branes as marginally deformed two-dimensional CFTs}

\author[a,b,1]{Suting~Zhao,\note[1]{These authors contributed equally to the project and share first authorship.}}
\author[c,1,2]{Gabriel Arenas-Henriquez,\note[2]{Corresponding author}}
\author[a,b,1]{Yizhou Lu,}
\author[d,a]{Hamed~Adami,}
\author[e,a]{Ren\'e Meyer}
\affiliation{$^a$ Shanghai Institute for Mathematics and Interdisciplinary Sciences (SIMIS), Shanghai, 200433, China}
\affiliation{$^b$ Research Institute of Intelligent Complex Systems, Fudan University, Shanghai 200433, China}
\affiliation{$^c$ School of Fundamental Physics and Mathematical Sciences, Hangzhou Institute for Advanced Study, UCAS, Hangzhou 310024, China}
\affiliation{$^d$ Center for Mathematics and Interdisciplinary Sciences, Fudan University, Shanghai, 200433, China}
\affiliation{$^e$Institute for Theoretical Physics and Astrophysics and W\"urzburg-Dresden Cluster of Excellence
ct.qmat, Julius-Maximilians-Universit\"at W\"urzburg, Am Hubland, 97074 W\"urzburg, Germany}

\emailAdd{suting.zhao@simis.cn, gabriel.arenas.henriquez@gmail.com, luyz@simis.cn, hadami@simis.cn, rene.meyer@uni-wuerzburg.de 
}

\abstract{We study how boundary correlation functions encode the charge dependence of the renormalised Euclidean action of an AdS$_3$ black brane. In three-dimensional Einstein--Maxwell theory, the Maxwell field has logarithmic asymptotics and is locally dual to a massless pseudoscalar. In the canonical ensemble, the charge density fixes the slope of a spatially linear source for the dual marginal operator. At fixed temperature, the classical action is even in the charge density due to reflection symmetry $\Phi\to-\Phi$, and its charge expansion coefficients are governed by Catalan numbers. We compute the quadratic and quartic terms from connected correlation functions around the neutral brane. The scalar two-point function and the shift symmetry of its source determine the matter Weyl anomaly, which fixes the logarithmic temperature dependence of the quadratic term in the charge expansion. At quartic order, the connected four-point function comes from tree-level graviton exchange. Integrating it against the linear source, with an infrared cutoff on the overall scale of the integral, reproduces the quartic coefficient of the action per unit length. At each higher order, the corresponding Catalan coefficient predicts the integrated sum of all connected tree-level Witten diagrams. 
}\maketitle

\section{Introduction}\label{sec:Intro}

The AdS/CFT correspondence relates gravitational dynamics in asymptotically anti-de Sitter spacetimes to observables in a conformal field theory at the boundary \cite{Maldacena:1997re,Gubser:1998bc,Witten:1998qj}. In top-down realisations, the microscopic boundary theory can often be specified explicitly as a quantum field theory. The standard example is the correspondence between four-dimensional $\mathcal{N}=4$ supersymmetric Yang--Mills theory and type IIB string theory on $\mathrm{AdS}_{5}\times S^{5}$. The situation is less direct in a bottom-up approach. A bulk effective action specifies a set of fields and interactions, while the choice of boundary conditions defines the associated variational problem. Together, these ingredients constrain a sector of the dual CFT in the regime where the bulk effective description is reliable \cite{Heemskerk:2009pn}. Identifying a microscopic theory that realises this CFT is a further part of the holographic problem.


At leading semiclassical order, the GKPW prescription identifies the renormalised bulk on-shell action with the generating functional of connected boundary correlation functions \cite{Gubser:1998bc,Witten:1998qj}. An exact classical solution therefore packages an infinite sequence of perturbative contributions into a single nonlinear expression. When the solution belongs to a family controlled by e.g. a charge or a source amplitude, its action can be expanded in that parameter, and the resulting coefficients can be compared with integrated connected correlators around a reference state. Our aim is to make this organisation explicit: how do these correlators combine to reproduce the charge dependence of a classical gravitational saddle? Such a comparison relates the thermodynamic description of the charged geometry to the interactions encoded in the boundary generating functional.

Einstein--Maxwell theory provides a natural setting for this question. In four and higher dimensions, electrically charged black holes have long served as examples in which gravitational thermodynamics can be studied in different ensembles. Fixing the electric charge or its conjugate potential requires different boundary terms in the Euclidean action. The importance of this distinction was emphasised by Hawking and Ross in their analysis of electric--magnetic duality and the semiclassical thermodynamics of charged black holes \cite{Hawking_1995}. In asymptotically AdS spacetimes, charged black holes exhibit a rich thermodynamic phase structure with a corresponding interpretation in the boundary theory \cite{Chamblin_1999}. With the usual Dirichlet boundary conditions, the bulk gauge field sources a conserved current associated with a global $U(1)$ symmetry, and the electric charge and chemical potential become the corresponding thermodynamic variables. More general boundary conditions lead to different boundary descriptions and deformations \cite{Marolf:2006nd}.

In three dimensions, Maxwell theory has several distinctive features. The gauge field carries a local propagating degree of freedom, and an electric charge produces logarithmic radial behaviour in both the gauge potential and the geometry. The resulting asymptotic structure makes the boundary conditions and renormalisation scheme an essential part of the holographic dictionary \cite{Marolf:2006nd,Jensen:2010em}. Although the Maxwell action is quadratic in the field strength, its stress tensor backreacts on the metric, generating nonlinear charge dependence in the thermodynamic quantities and the on-shell action. Charged BTZ geometries consequently offer a simple setting in which propagating matter, gravitational backreaction and logarithmic asymptotics can be studied together. Their finite-density dynamics were explored by Faulkner and Iqbal \cite{Faulkner:2012gt}, who used the scalar dual of the Maxwell field to analyse density correlations and the Friedel oscillations generated by monopole effects.

The Maxwell description differs from the familiar Chern--Simons realisation of charged AdS$_3$ holography. Pure Chern--Simons theory has no local bulk degrees of freedom, while its standard chiral boundary conditions give rise to an affine current algebra (see for instance \cite{Kraus:2006wn} and reference therein). This structure has proved particularly useful for symmetry-resolved entanglement, where the reduced density matrix is decomposed into sectors of fixed charge. In the $U(1)$ Chern--Simons setting, charged moments admit a description in terms of bulk Wilson lines \cite{Belin:2013uta} and boundary vertex operators, allowing direct comparisons between holographic and CFT calculations \cite{Zhao:2020qmn,Zhao:2024ugx}. For Maxwell theory, the logarithmic modes and the choice of boundary conditions lead to a different organisation of the boundary data. In the fixed-flux formulation used below, the electric momentum is prescribed, and the local response can be expressed through the operator dual to a massless bulk scalar \cite{Jensen:2010em,Faulkner:2012gt}.

Entanglement observables provide further motivation for understanding these distinctions. In three-dimensional compact Maxwell theory, the scalar dual makes the contribution of winding sectors to the disk entanglement and R\'enyi entropies explicit \cite{Agon:2013iva}. In four dimensions, comparisons between Maxwell and scalar entanglement reveal the importance of the allowed mode content \cite{Casini:2015dsg}. Holographically, charged R\'enyi entropies establish another connection between gauge fields, entanglement and black-hole thermodynamics: introducing a chemical potential for the charge relates these observables to the thermodynamics of charged black holes with hyperbolic horizons \cite{Belin:2013uta,Bueno:2022jbl,Arenas-Henriquez:2022ntz}. These developments motivate a careful treatment of the charge dependence of gravitational generating functionals and of the boundary data that define them.

In this work, we study three-dimensional Einstein gravity with a negative cosmological constant  coupled to a Maxwell field. Aspects of holography of this setup has already been discussed in \cite{Marolf:2006nd,Jensen:2010em}. We focus on the electrically charged black brane with a non-compact spatial direction and work in the canonical ensemble. The boundary terms fix the electric flux, and holographic renormalisation gives a finite Euclidean action satisfying the canonical quantum statistical relation. At leading semiclassical order, the Maxwell field can be described locally by a massless scalar, a description, which was heavily used in \cite{Jensen:2010em} The corresponding first-order formulation maps the fixed-flux variational problem to scalar Dirichlet boundary conditions and keeps track of the boundary term required for the equality of the saddle actions.

In the scalar description, the electric charge density has a simple boundary interpretation. In standard quantisation, a massless scalar in AdS$_3$ is dual to an operator $\mathcal O_\phi$ of dimension two. Its boundary value $\phi$ is a dimensionless source that enters the CFT action through
\begin{equation}
    S_{\mathrm{CFT}}\longrightarrow S_{\mathrm{CFT}}
    +\int_\Sigma d^2x\,\sqrt{\gamma}\,\phi\,\mathcal O_\phi\,.
\end{equation}
For the charged black brane, the scalar profile is
\begin{equation}
    \Phi(r,x)=\phi(x)=-e^2qx\,,
\end{equation}
where $q$ is the electric charge density. A spatial translation shifts $\phi$ by a constant. Since the bulk action depends only on derivatives of $\Phi$, this spatially varying marginal source is compatible with a homogeneous geometry.

Varying $q$ at fixed temperature changes the slope of the source. Expanding the generating functional in this slope therefore relates the charge dependence of the brane action to connected scalar correlators in the neutral thermal state. The first nonlinear term raises a simple question. The Maxwell action is quadratic in the field strength, yet the free energy contains a term of order $q^4$. In the charged solution this term arises from the backreaction of the Maxwell field on the geometry. In perturbation theory around the neutral brane, the same interaction appears as graviton exchange in the connected scalar four-point function.

We compare these two descriptions through order $q^4$. The two-point function sets the normalisation and determines the matter Weyl anomaly. The four-point function then provides the first test of gravitational backreaction. Beyond this order, the Catalan coefficients of the exact brane action
predict the integrated connected higher-point functions. In bulk perturbation theory, the coefficient of $q^{2n}$ encodes the sum of all connected tree-level Witten diagrams with $2n$ external scalar legs, evaluated around the neutral brane and integrated against the linear source.

The paper is organised as follows. In \autoref{sec:einstein-maxwell} we describe the variational principle, the scalar representation and holographic renormalisation. In \autoref{sec:charged-brane} we study the charged solution and its semiclassical charge expansion. In \autoref{sec:dual-CFT} we evaluate the quadratic and quartic source contributions. We conclude in \autoref{sec:discussion} with the limitations of the comparison and possible extensions.

\section{Einstein--Maxwell theory in \texorpdfstring{AdS$_3$}{AdS3}}
\label{sec:einstein-maxwell}

We consider three--dimensional Euclidean Einstein--Maxwell theory with a negative cosmological constant. The bulk dynamics is governed by the Einstein--Hilbert action supplemented by the cosmological constant term and the Maxwell contribution. The corresponding Euclidean action takes the form
\begin{equation}
    I_{\mathrm E}
    =
    -\frac{1}{2\kappa^2}
    \int_{\mathcal M} d^3x\sqrt{g}
    \left(R+\frac{2}{\ell^2}\right)
    +\frac{1}{4e^2}  \int_{\mathcal M}d^3x\, \sqrt{g} \, F_{\mu\nu}F^{\mu\nu}+I_{\mathrm{bdy}}
    +I_{\mathrm{ct}} \, ,
    \label{eq:EM-action-new}
\end{equation}
where $\mathcal M$ denotes the three-dimensional spacetime manifold, $g_{\mu\nu}$ is the Euclidean metric, $R$ is its Ricci scalar, and $\ell$ is the AdS radius. The gravitational coupling is denoted by $\kappa^2$, while $e$ is the Maxwell coupling. The electromagnetic field strength is $F_{\mu\nu}=\partial_\mu A_\nu-\partial_\nu A_\mu$, where $A_\mu$ is the gauge potential. We denote the sum of the two bulk integrals by $I_{\rm bulk}$. The terms $I_{\mathrm{bdy}}$ and $I_{\mathrm{ct}}$ represent boundary contributions and counterterms, respectively. Their precise forms depend on the boundary conditions and variational principle adopted for the metric and gauge field, as well as on the renormalisation prescription.

The negative cosmological constant is $\Lambda=-1/\ell^2$, so the theory admits asymptotically AdS$_3$ configurations. The Maxwell field modifies the gravitational dynamics through its stress-energy tensor and can also introduce nontrivial boundary data associated with the gauge sector. In three dimensions, these effects must be treated with particular care because the Maxwell field can generate logarithmic radial behaviour in asymptotically AdS solutions. Consequently, the specification of boundary conditions and the associated boundary terms is essential for defining a well-posed variational problem and a consistent set of boundary observables.

The bulk dynamics is governed by the Einstein and Maxwell equations,
\begin{subequations}
    \begin{align}
        R_{\mu\nu}
    -\frac{1}{2}\, R\,g_{\mu\nu}
    -\frac{1}{\ell^2}\, g_{\mu\nu}
    &=
    \frac{\kappa^2}{e^2}
    \left(
        F_{\mu\rho}F_\nu{}^\rho
        -\frac{1}{4}g_{\mu\nu}F_{\rho\sigma}F^{\rho\sigma}
    \right),
    \label{eq:einstein-eom-new}\\
    \nabla_\mu F^{\mu\nu}&=0 \,  .
    \label{eq:maxwell-eom-new}
    \end{align}
\end{subequations}
Taking the trace of the Einstein equation yields a direct relation between the Ricci scalar and the Maxwell invariant,
\begin{equation}
    R
    =
    -\frac{6}{\ell^2}
    -\frac{\kappa^2}{2e^2}
    F_{\mu\nu}F^{\mu\nu}.
    \label{eq:trace-einstein-new}
\end{equation}
This expression shows that the electromagnetic field generally modifies the scalar curvature away from its vacuum value, $R=-6/\ell^2$. In particular, the Ricci scalar need not be constant when the Maxwell invariant varies throughout spacetime.

In three dimensions, the Maxwell field strength admits a dual description in terms of a vector field,
\footnote{We adopt the convention $F_{\mu\nu}=\epsilon_{\mu\nu\lambda}\hat{F}^{\lambda}$.}
\begin{equation}
    \hat{F}^\mu= \frac{1}{2}\, \epsilon^{\mu\alpha\beta}\, F_{\alpha\beta} \, ,
\end{equation}
where $\epsilon^{\mu\alpha\beta}$ is the Levi-Civita tensor. Under this duality, the Bianchi identity for the Maxwell field is equivalent to the divergence-free condition $\nabla_\mu\hat{F}^{\mu}=0$, whereas the source-free Maxwell equation \eqref{eq:maxwell-eom-new} implies the closure condition $\nabla_{[\mu}\hat{F}_{\nu]}=0$. The latter condition guarantees that the dual one-form is locally exact, so that the dual vector can be expressed in terms of a scalar potential. Consequently, in a simply connected patch, one may introduce a scalar potential $\Phi$ such that
\begin{equation}
    \hat{F}_\mu= -i\, \partial_\mu \Phi \, ,
\end{equation}
where the factor of $i$ accounts for the chosen continuation of a real Lorentzian Maxwell field to Euclidean signature. The divergence-free condition then reduces to the massless scalar equation
\begin{equation}\label{eq:scalar}
    \square \Phi =0 \, .
\end{equation}
This scalar description is local: on a spacetime with nontrivial topology, a closed dual one-form need not be globally exact, and additional global data may therefore be required.

Substituting the dual scalar representation into the Einstein equation \eqref{eq:einstein-eom-new} gives
\begin{equation}\label{eq:eom-Einstein-Scalar}
    R_{\mu\nu} =-\frac{2}{\ell^2} \,g_{\mu\nu}+
    \frac{\kappa^2}{e^2}\,
        \partial_\mu \Phi \, \partial_\nu \Phi \, .
\end{equation}
Thus, in three dimensions, the coupled Einstein--Maxwell system can be expressed locally as Einstein gravity coupled to a massless scalar field, subject to the stated duality and signature conventions. The scalar potential is defined only up to an additive constant,
\begin{equation}
    \Phi \quad \longrightarrow \quad \Phi +\lambda \, ,
\end{equation}
where $\lambda$ is a constant parameter. This freedom leaves $\hat{F}_\mu$ and hence the original Maxwell field strength unchanged.

The relations above establish a local equivalence at the level of the classical equations of motion. Extending this correspondence to the quantum theory requires additional information about the global sectors and the operator content. Related subtleties arise in four-dimensional comparisons between Maxwell and scalar fields, where differences in the allowed modes lead to distinct entanglement observables \cite{Casini:2015dsg,Abate:2025ywp}. Furthermore, as emphasised in \cite{Witten:2026twr}, quantum duality in Euclidean signature may require Poisson resummation over quantised flux sectors, and a semiclassical description in one formulation need not correspond to a semiclassical description in the other. 

Here, we use the scalar formulation at leading semiclassical order for the classical saddles under consideration. This requires supplementing the matching of the bulk equations with a consistent identification of the variational principles, boundary terms, and boundary conditions in the chosen ensemble. We therefore introduce a first-order action in which the field strength and the scalar are independent fields, and derive the scalar description together with the boundary term required by the Maxwell variational principle.

\subsection{Variational principle}\label{sec:Vari}
Let $z$ be a radial defining function for the conformal boundary $\Sigma$, with $z=0$ on $\Sigma$ and $z>0$ in the bulk. The Fefferman--Graham form of the metric and its asymptotic expansion will be specified below. We regulate the bulk by introducing a cutoff surface $\Sigma_\varepsilon$ at $z=\varepsilon$, and denote the retained region by $\mathcal M_\varepsilon$. We assume that this surface is smooth and constitutes the only boundary component, with no edges:
\begin{equation*}
    \partial\mathcal M_\varepsilon=\Sigma_\varepsilon,
    \qquad
    \partial\Sigma_\varepsilon=\varnothing.
\end{equation*}
We keep the boundary embedding fixed and denote its outward unit normal by $n^\mu$. The induced metric and extrinsic curvature are
\begin{equation*}
    \mathrm g_{\mu\nu}=g_{\mu\nu}-n_\mu n_\nu,
    \qquad
    K_{\mu\nu}
    =\mathrm g_\mu{}^\alpha\mathrm g_\nu{}^\beta
    \nabla_\alpha n_\beta.
\end{equation*}
For a non-compact boundary, we take variations of compact support in the boundary directions when deriving the local variational identities, so that tangential integrations by parts produce no contributions from spatial infinity.

We first choose the boundary terms that specify the variational problem. For the metric, the Gibbons--Hawking--York term removes normal derivatives of its variation. At fixed metric, the Maxwell variation of the bulk action is
\begin{equation}
\delta_A I_{\rm bulk}\big|_{\rm EOM}
=\int_{\Sigma_\varepsilon}d^2x\,\Pi^a\delta A_a,
\qquad
\Pi^a:=\frac{\sqrt{\mathrm g}}{e^2}\,n_\mu F^{\mu a}.
\end{equation}
To fix the electric momentum density $\Pi^a$, we include the Maxwell Legendre term. The complete boundary contribution is
\begin{equation}\label{eq:bdy-canonical-new}
I_{\rm bdy}
=-\frac{1}{\kappa^2}\int_{\Sigma_\varepsilon}d^2x\sqrt{\mathrm g}\,K
-\int_{\Sigma_\varepsilon}d^2x\,A_a\Pi^a.
\end{equation}
We define the regulated action at finite cutoff by
\begin{equation}\label{eq:regulated-action-definition}
I_{\rm reg}(\varepsilon)
:=I_{\rm bulk}[\mathcal M_\varepsilon]+I_{\rm bdy}[\Sigma_\varepsilon].
\end{equation}
This is an action functional before imposing the field equations and does not include counterterms. Its Maxwell variation at fixed metric is
\begin{equation}
\delta_A I_{\rm reg}\big|_{\rm EOM}
=-\int_{\Sigma_\varepsilon}d^2x\,A_a\delta\Pi^a.
\end{equation}
Thus the variational principle fixes the electric flux, corresponding to the canonical ensemble for the charged brane considered below. Maxwell boundary conditions in AdS$_3$ and their scalar description are discussed in \cite{Jensen:2010em,Faulkner:2012gt}.

To derive the scalar representation of this variational problem, we use a first-order formulation. We introduce an independent two-form $F$ and a scalar $\Phi$, with first-order action
\begin{equation}\label{eq:first-order-maxwell}
    I_{\rm first}
    =\frac{1}{2e^2}\int_{\mathcal M_\varepsilon}F\wedge *F
    +\frac{i}{e^2}\int_{\mathcal M_\varepsilon}d\Phi\wedge F.
\end{equation}
Variation with respect to $F$ reproduces the duality relation,
\begin{equation}\label{eq:Fscalar}
    *F+i\,d\Phi=0,
    \qquad
    F=-i*d\Phi,
\end{equation}
where we used $*^2=1$ on one- and two-forms in Euclidean three dimensions. Variation with respect to $\Phi$ gives
\begin{equation}
    \delta_\Phi I_{\rm first}
    =-\frac{i}{e^2}\int_{\mathcal M_\varepsilon}\delta\Phi\,dF
    +\frac{i}{e^2}\int_{\Sigma_\varepsilon}\delta\Phi\,F.
\end{equation}
With Dirichlet boundary conditions,
$\delta\Phi|_{\Sigma_\varepsilon}=0$, this imposes the Bianchi identity $dF=0$, which locally implies $F=dA$. The mixed term in \eqref{eq:first-order-maxwell} then becomes
\begin{equation}
    \frac{i}{e^2}\int_{\mathcal M_\varepsilon}d\Phi\wedge F
    =\frac{i}{e^2}\int_{\Sigma_\varepsilon}A\wedge d\Phi.
\end{equation}

To identify this boundary contribution, choose coordinates
$x^a=(\tau,x)$ and bulk orientation
$dz\wedge d\tau\wedge dx>0$. Using \eqref{eq:Fscalar}, the electric momentum at the cutoff surface is
\begin{equation}\label{eq:flux-scalar-source}
    \Pi^a
    =\frac{i}{e^2}\varepsilon^{ab}
    \partial_b\Phi\big|_{\Sigma_\varepsilon},
    \qquad
    \varepsilon^{\tau x}=+1,
\end{equation}
where $\varepsilon^{ab}$ is the antisymmetric symbol. Since the outward normal points towards decreasing $z$, the induced orientation on $\Sigma_\varepsilon$ gives
\begin{equation}
    \frac{i}{e^2}\int_{\Sigma_\varepsilon}A\wedge d\Phi
    =-\frac{i}{e^2}\int d\tau\,dx\,
    A_a\varepsilon^{ab}\partial_b\Phi
    =-\int_{\Sigma_\varepsilon}d^2x\,A_a\Pi^a.
\end{equation}
The first-order action therefore reproduces precisely the Maxwell boundary term in \eqref{eq:bdy-canonical-new}. This identification uses the duality relation and the absence of contributions from patch boundaries or spatial infinity. On the non-compact brane, the action-level identification is understood with the infrared prescription used for the homogeneous action density; compact support of variations alone does not establish it. 
Alternatively, eliminating $F$ directly from \eqref{eq:first-order-maxwell} yields
\begin{equation}
    \left.I_{\rm first}\right|_{F=-i*d\Phi}
    =-\frac{1}{2e^2}\int_{\mathcal M_\varepsilon}d\Phi\wedge *d\Phi
    +\frac{1}{e^2}\int_{\mathcal M_\varepsilon}d\Phi\wedge *d\Phi
    =\frac{1}{2e^2}\int_{\mathcal M_\varepsilon}d\Phi\wedge *d\Phi.
\end{equation}
Both terms in the first-order action are essential to obtain the positive scalar kinetic term.

The relation between the boundary conditions follows from \eqref{eq:flux-scalar-source}. The Maxwell constraint is automatically satisfied,
\begin{equation}\label{eq:radial-momentum-scalar}
    \partial_a\Pi^a
    =\frac{i}{e^2}\varepsilon^{ab}
    \partial_a\partial_b\Phi\big|_{\Sigma_\varepsilon}
    =0,
\end{equation}
and, conversely, a conserved electric momentum locally determines the boundary scalar up to an additive constant. Fixing $\Pi^a$ therefore fixes $\Phi|_{\Sigma_\varepsilon}$ modulo this constant shift. After choosing a representative, this is the scalar Dirichlet problem used above. In the boundary limit, writing $\Phi\to\phi$, the relation becomes
\begin{equation}
    \lim_{\varepsilon\to0}\Pi^a
    =\frac{i}{e^2}\varepsilon^{ab}\partial_b\phi.
\end{equation}
The freedom $\phi\to\phi+\lambda$ leaves the prescribed electric flux unchanged.

Restoring gravity, the regulated action in the scalar formulation is
\begin{equation}\label{eq:regulated-scalar-action}
    \begin{aligned}
I_{\rm reg}[g,\Phi;\varepsilon]={}&
-\frac1{2\kappa^2}\int_{\mathcal M_\varepsilon}d^3x\sqrt g
\left(R+\frac2{\ell^2}\right)
+\frac1{2e^2}\int_{\mathcal M_\varepsilon}d^3x\sqrt g\,
\nabla_\mu\Phi\nabla^\mu\Phi\\
&-\frac1{\kappa^2}\int_{\Sigma_\varepsilon}d^2x
\sqrt{\mathrm g}\,K.
\end{aligned}
\end{equation}

In Maxwell variables, the on-shell variation at fixed cutoff contains $-\int A_a\delta\Pi^a$. Substituting \eqref{eq:flux-scalar-source} and integrating by parts along $\Sigma_\varepsilon$ gives
\begin{equation}\label{variation-Lagrangian-on-shell}
\begin{aligned}
\left.\delta I_{\rm reg}\right|_{\rm on\text{-}shell}
=&\int_{\Sigma_\varepsilon}d^2x\sqrt{\mathrm g}
\left[\frac{1}{2\kappa^2}
\left(K^{\alpha\beta}-K\mathrm g^{\alpha\beta}\right)
\delta \mathrm{g}_{\alpha\beta}
+\frac{1}{e^2}\,n\cdot\nabla\Phi\,\delta\Phi
\right]\\
&-\int_{\Sigma_\varepsilon}d\Sigma_\mu
\nabla_\nu\!\left(
\frac{1}{\kappa^2}\,n^{[\mu}\delta n^{\nu]}
+\frac{i}{e^2}
\epsilon^{\mu\nu\lambda}A_\lambda\delta\Phi
\right).
\end{aligned}
\end{equation}
Here $d\Sigma_\mu=n_\mu\sqrt{\mathrm g}\,d^2x$. Direct variation of the scalar action \eqref{eq:regulated-scalar-action} gives the same expression with the $A_\lambda\delta\Phi$ term omitted. This term is the edge contribution from the Maxwell integration by parts. Both total derivatives integrate to zero on a closed boundary or for compactly supported variations.

We choose Dirichlet boundary conditions for the induced metric and the scalar,
\begin{equation}
    \delta\mathrm g_{ab}\big|_{\Sigma_\varepsilon}=0,
    \qquad
    \delta\Phi\big|_{\Sigma_\varepsilon}=0.
\end{equation}
Under these conditions, the on-shell variation vanishes and the regulated action defines a stationary variational problem. As shown above, fixing the boundary scalar fixes the Maxwell electric momentum; conversely, the electric momentum determines the scalar locally up to an additive constant.

After renormalisation and removal of the cutoff, the corresponding boundary conditions fix the leading metric and scalar data,
\begin{equation}
    \delta\gamma_{ab}=0,
    \qquad
    \delta\phi=0.
\end{equation}
These conditions specify the boundary-value problem for each prescribed pair of sources $(\gamma_{ab},\phi)$. To compute the holographic one-point functions, we instead differentiate the renormalised on-shell action with respect to these sources, comparing solutions with different boundary data.

\subsection{Asymptotic expansion}\label{sec:AsymExp}

We now choose Fefferman--Graham gauge in which the bulk metric takes the form
\begin{equation}\label{eq:FG}
    d s^{2}=\frac{\ell^2}{z^{2}} d z^{2}+\mathrm{g}_{ab}(z,x)dx^{a}dx^{b}\,.
\end{equation}
The conformal boundary is located at $z=0$. Choosing the unit normal to point outward, from the bulk region towards the conformal boundary, we have
\begin{equation}\label{eq:normal-new}
    n^\mu\partial_\mu=-\frac{z}{\ell}\, \partial_z \, ,
\qquad
n_\mu dx^\mu=-\frac{\ell}{z}\, \d z\,.
\end{equation}
With the induced metric and extrinsic-curvature conventions introduced above, the nonvanishing components are
\begin{equation}\label{eq:induced-extrinsic-new}
K_{ab}=-\frac{z}{2\ell}\,\partial_z\mathrm{g}_{ab}\,.
\end{equation}

Near the conformal boundary, we expand the induced metric in powers of $z$, allowing for logarithmic terms required by the coupled field equations. Solving the radial evolution equations to the relevant orders (see Appendix~\ref{app:holo-ren}) gives
\begin{equation}\label{eq:asymptotic-metric-new}
    \begin{split}
        \mathrm{g}_{ab}(z,x)
 = &
\frac{\ell^2}{z^2}\, \gamma_{ab}
+
\kappa^2\ell \, T_{ab}
-\gamma_{ab} 
\left( \frac{\ell^2}{2} \,\mathcal R[\gamma]
- \frac{\kappa^2\ell^2}{2e^2}\,
D_a \phi D^a \phi\right)\\
&+
\frac{\kappa^2\ell^2}{2e^2}
D_{\langle a}\phi
D_{b\rangle}\phi \left( 1-
\ln\frac{z^2}{\ell^2}\right)
+
O\!\left(\frac{z^2}{\ell^2}\ln^2\left|\frac{z}{\ell}\right|\right) \,.
    \end{split}
\end{equation}
Here, $\gamma_{ab}(x)$ is the boundary metric, $\phi(x)$ is the leading boundary value of the scalar field, and $T_{ab}(x)$ parametrises the finite, state-dependent contribution to the metric expansion. Also, $\mathcal R[\gamma]$ denotes the Ricci scalar of the boundary metric, and $D_a$ is the covariant derivative compatible with $\gamma_{ab}$. Boundary indices are raised and lowered using $\gamma_{ab}$. The angular brackets denote the symmetric, traceless projection with respect to $\gamma_{ab}$. The logarithmic term reflects the backreaction of the scalar field and is part of the asymptotic solution; it cannot in general be omitted when computing the subleading geometric data.

We similarly expand the bulk scalar field as
\begin{equation}\label{eq:asymptotic-phi-new}
   \Phi(z,x)=\phi + \frac{z^2}{\ell^2}\left[-\frac{e^2\ell}{2}\psi\, 
+\frac{\ell^2}{4}
D^2\phi \left( 1-\ln \frac{z^{2}}{\ell^2}\right)\right] + \mathcal{O}\!\left(\frac{z^4}{\ell^4}\ln^2\left|\frac{z}{\ell}\right|\right)\,.
\end{equation}
where $\psi=\psi(x)$ denotes the independent subleading scalar datum in this expansion, while $D^2:=D_aD^a$ is the Laplacian associated with $\gamma_{ab}$. The leading coefficient $\phi$ is the Dirichlet scalar source selected above. We have parametrised the independent finite data by $T_{ab}$ and $\psi$ so that they coincide with the renormalised responses in the reference scheme. Appendix~\ref{app:holo-ren} derives this parametrisation from the radial equations and the canonical momenta.

The bulk equations of motion impose constraints on the asymptotic coefficients. In particular, defining the trace $T:=\gamma^{ab}T_{ab}$, the trace equation \eqref{eq:trace-einstein-new} leads to
\begin{equation}\label{eq:Weyl-anomaly-new}
T
=
\frac{c}{24\pi}
\mathcal R[\gamma]
-
\frac{\ell}{2e^2}
D_a\phi D^a\phi\,,
\end{equation}
where $c=12\pi\ell/\kappa^{2}$ is the Brown--Henneaux central charge. This equation fixes the trace of $T_{ab}$ in terms of the boundary geometry and scalar source, leaving its traceless part subject to the remaining equations and boundary data. The curvature contribution is the familiar geometric term associated with the two-dimensional trace anomaly, while the scalar-gradient term describes the modification induced by the boundary scalar.

The mixed components of the bulk equations also yield a constraint on the divergence of $T_{ab}$,
\begin{equation}\label{eq:div-constraint}
    D^bT_{ab}= -\psi\,  \partial_{a}\phi\,.
\end{equation}
This equation expresses the failure of $T_{ab}$ to be separately conserved in the presence of a nonconstant scalar source: its divergence is governed by the scalar datum $\psi$ and the gradient of $\phi$. In particular, for a constant boundary scalar source, the constraint reduces to covariant conservation of $T_{ab}$. Together, the trace and divergence relations constrain the asymptotic data and provide consistency conditions for the boundary stress tensor and scalar sector.

\subsection{Renormalised action and one-point functions}\label{subsec:holo-ren-onepoint}
The remaining task is to remove the asymptotic divergences in the on-shell value of \eqref{eq:regulated-scalar-action} and its variation.
In the scalar formulation, the holographic renormalisation of the canonical Einstein--Maxwell theory reduces to the standard problem of Einstein gravity coupled to a massless scalar. This procedure has been extensively studied, including considerably more general scalar couplings and boundary conditions \cite{deHaro:2000xn,Skenderis:2002wp,Papadimitriou:2004ap,Papadimitriou:2007sj, Papadimitriou:2010as,Jensen:2010em,Anabalon:2015xvl,Caceres:2023gfa}. Scalar backreaction and logarithmic branches in asymptotically AdS spacetimes were analysed in \cite{Henneaux:2004zi}; three-dimensional examples with scalar hair and non-trivial boundary contributions were studied in \cite{Cisterna:2023qhh}. For the particular massless scalar arising from Maxwell duality in AdS$_{3}$, the explicit cancellation of divergences and derivation of the one-point functions is spelled out in \autoref{app:holo-ren}.

Evaluating the regulated action on solutions with the Fefferman--Graham expansion above gives its divergent terms
\begin{equation}\label{eq:reg-div-new}
\begin{aligned}
I_{\rm reg}^{\rm div}={}&-\frac{\ell}{\kappa^2\varepsilon^2}
\int_{\Sigma}d{}^2x\sqrt\gamma+\frac{\ell}{2\kappa^2}\log\frac{\varepsilon}{\ell}
\int_{\Sigma}d{}^2x\sqrt\gamma\,\mathcal R[\gamma]\\
&-\frac{\ell}{2e^2}\log\frac{\varepsilon}{\ell}
\int_{\Sigma}d{}^2x\sqrt\gamma\,\partial _a\phi \partial ^a\phi,
\end{aligned}
\end{equation}
A local counterterm that cancels these divergences is
\begin{equation}\label{eq:counterterm}
\begin{aligned}
    I_{\rm ct}=&\frac{1}{\kappa^2 \ell}\int_{\Sigma_{\varepsilon}}d^2 x \sqrt{\rm g}-\frac{\ell}{2\kappa^2} \log \frac{\varepsilon}{\ell}
\int_{\Sigma_{\varepsilon}}d^2 x \sqrt{\mathrm{g}} \mathcal{R}[\mathrm{g}]\\
&+\frac{\ell}{2e^2}\log \frac{\varepsilon}{\ell} \int_{\Sigma_{\varepsilon}}d^2 x \sqrt{\rm g}\mathrm{g}^{ab} \partial _a \Phi \partial _b \Phi+ I_{\rm fin}.
\end{aligned}
\end{equation}
The first two terms correspond to the leading volume counterterm and the curvature contribution to the logarithmic term are the standard gravitational counterterms in three dimensions \cite{Balasubramanian:1999re,deHaro:2000xn, Miskovic:2006tm}. The remaining logarithmic contribution is the scalar counterterm associated with the marginal source \cite{deHaro:2000xn,Jensen:2010em}.  Here $I_{\rm fin}[\gamma,\phi]$ is a finite local functional of the sources specifying the renormalisation scheme. All cutoff contractions in this expression use $\mathrm g_{ab}$ and the bulk scalar $\Phi$. 
We define
\begin{equation}
I_{\rm ren}:=\lim_{\varepsilon\to0}\bigl(I_{\rm reg}(\varepsilon)+I_{\rm ct}(\varepsilon)\bigr).
\end{equation}
Combining the regulated variation with the counterterm variation and taking the cutoff to zero gives
\begin{equation}\label{eq:ren-Onshell}
    \delta I_{\rm ren}|_{\rm on-shell}=\int _{\Sigma}\mathrm{d} ^2 x \sqrt{\gamma} \left(
-\frac{1}{2}T^{ab}\, \delta \gamma_{ab}+\psi\, \delta \phi 
\right)+\delta I_{\rm fin}.
\end{equation}
In the reference scheme $I_{\rm fin}=0$, this identifies $\langle T^{ab}\rangle=T^{ab}$ and $\langle\mathcal O_\phi\rangle=\psi$. A finite local term shifts them to
\begin{align}\label{eq:T-finite}
\langle T^{ab}\rangle&=T^{ab}-\frac{2}{\sqrt\gamma}\frac{\delta I_{\rm fin}}{\delta\gamma_{ab}},\\
\label{eq:psi-finite}
\langle\mathcal O_\phi\rangle&=\psi+\frac{1}{\sqrt\gamma}\frac{\delta I_{\rm fin}}{\delta\phi}.
\end{align}
The trace and divergence constraints of the preceding subsection are therefore the Ward identities in the reference scheme, which we use below. 
The source variations in \eqref{eq:ren-Onshell} define $T^{ab}$ and $\psi$ by comparing nearby boundary data.
For the Dirichlet boundary-value problem itself, we instead impose $\delta\gamma_{ab}=0$ and $\delta\phi=0$. The renormalised on-shell action is then stationary, with edge contributions absent under the boundary and infrared conditions stated above.

\subsection{Further comments on the bulk setup}\label{sec:Holography}

\paragraph{Fixed electric flux and the scalar source.}
The variational principle of \autoref{sec:Vari} provides a scalar description of the canonical Maxwell ensemble. Locally, the duality $F=-i*d\Phi$ identifies the electric flux with the tangential gradient of the scalar. 
Fixing that flux therefore fixes the boundary source $\phi$ up to an additive constant. 
Choosing this constant gives the Dirichlet scalar problem used here. The boundary condition fixes the local electric momentum density $\Pi^a(x)$. On the homogeneous saddle this fixes the total charge $Q=2Wq$; fixing only $Q$ would also allow spatial variations of the flux whose integral vanishes. 
The bulk field $\Phi$ retains the single local propagating degree of freedom of 3d Maxwell theory.\footnote{The dynamical role of this mode is particularly explicit for asymptotically flat boundary conditions: electromagnetic radiation can carry energy to null infinity, producing news, charge fluxes and a Bondi mass-loss law \cite{Bosma:2023sxn}; see also the earlier asymptotic analysis of \cite{Barnich:2015jua}. This is a separate boundary problem from the asymptotically AdS one considered here.} In standard quantisation, this massless AdS$_3$ scalar is dual to a scalar operator $\mathcal O_\phi$ of dimension $\Delta=2$, with source $\phi$. Aspects of the scalar description of AdS$_3$ Maxwell theory and charged BTZ backgrounds were studied in \cite{Faulkner:2012gt}.

For the charged black brane below, the source is $\phi(x)=-e^2qx$, as in \eqref{eq:brane-scalar}. The electric charge density thus determines the spatial gradient of a source for a marginal operator. Evaluating the renormalised action on this background gives the generating functional at that source, which is the starting point for the subsequent expansion in integrated scalar correlators.

\paragraph{The logarithmic Maxwell data.}
The scalar source also determines the logarithmic mode of the gauge potential. In radial gauge $A_z=0$,
\begin{equation}\label{eq:maxwell-asymptotics}
A_a(z,x)=-\widetilde a_a(x)\log\frac z\ell+a_a(x)+\cdots,
\qquad
\widetilde a^{\,a}=i\ell\,\epsilon_\gamma^{ab}D_b\phi,
\end{equation}
where $\epsilon_\gamma^{ab}=\varepsilon^{ab}/\sqrt\gamma$ and boundary indices are raised with $\gamma_{ab}$. Indeed, the leading momentum is $\Pi^a=\sqrt\gamma\,\widetilde a^{\,a}/(e^2\ell)$, in agreement with \eqref{eq:flux-scalar-source}. A change of defining function $z'=e^\omega z$, with the compensating tangential transformation that preserves Fefferman--Graham gauge, gives
\begin{equation}
\gamma'_{ab}=e^{2\omega}\gamma_{ab},\qquad
\widetilde a'_a=\widetilde a_a,\qquad
a'_a=a_a+\omega\widetilde a_a
\end{equation}
at the displayed order, up to a boundary gauge transformation. Thus the finite gauge potential depends on the reference scale \cite{Marolf:2006nd,Jensen:2010em}. Fixing its appropriately renormalised value defines a different Maxwell variational problem. In the canonical problem used here, the gradient of $\phi$ instead fixes the electric flux.

\paragraph{On boundary conditions and asymptotic symmetries.}
The fixed-flux prescription should be distinguished from the boundary conditions used to classify integrable asymptotic charges in \cite{Perez:2015jxn}. In the class studied there, the leading electromagnetic modes are allowed to vary and are related by a boundary functional. On a cylindrical boundary, a generic choice preserves time translations, spatial rotations and the global electromagnetic $U(1)$.
A special one-parameter family preserves two Virasoro algebras with the Brown--Henneaux central extension, together with the global $U(1)$. 
These results concern the choice of phase space. The condition $\delta\Pi^a=0$ adopted here does not by itself identify our variational problem with the special conformal family or establish the same asymptotic symmetry algebra. 
Likewise, covariance of the generating functional under transformations of its sources is distinct from invariance of a fixed source profile.

The global $U(1)$ in that conformal family does not remove the local Maxwell mode described by $\Phi$.
Pure $U(1)$ Chern--Simons theory, by contrast, has no local bulk mode and can support an affine current algebra under suitable boundary conditions. 
The boundary charge algebra and the local bulk degree of freedom are distinct pieces of information: an affine current algebra is neither implied nor excluded solely by the presence of a Maxwell photon. 
Our use of the scalar generating functional follows from the fixed-flux variational principle and local duality, rather than from an identification with the special conformal boundary conditions.

\paragraph{Boundary responses and Ward identities.}
The renormalised action constructed in \autoref{subsec:holo-ren-onepoint} defines the boundary generating functional with sources $\gamma_{ab}$ and $\phi$. Its variation \eqref{eq:ren-Onshell} identifies $\langle T_{ab}\rangle=T_{ab}$ and $\langle\mathcal O_\phi\rangle=\psi$ in the reference scheme $I_{\rm fin}=0$. Thus the asymptotic data of \autoref{sec:AsymExp} acquire the interpretation of boundary one-point functions.

With this identification, the radial constraint \eqref{eq:div-constraint} is the diffeomorphism Ward identity: a non-constant scalar source exchanges momentum with the boundary theory. The trace relation \eqref{eq:Weyl-anomaly-new} is the holographic Weyl anomaly \cite{Henningson:1998gx,deHaro:2000xn}. Its curvature term carries the Brown--Henneaux central charge, while the scalar-gradient term is the contribution of the marginal source. The latter fixes the logarithmic part of the quadratic generating functional discussed below.

The constant shift symmetry $\phi\to\phi+\lambda$ supplies a further integrated identity:
\begin{equation}
\int d^2x\sqrt{\gamma}\,\langle\mathcal O_\phi\rangle=0\,.
\end{equation}
This follows by applying \eqref{eq:ren-Onshell} to a constant scalar variation at fixed metric, provided that the shift produces no boundary contribution. 

\paragraph{Finite scheme dependence.}
The finite-response formulas \eqref{eq:T-finite} and \eqref{eq:psi-finite} describe how local counterterms change this identification. We set finite source functionals with more than two boundary derivatives to zero; a term proportional to $\int\sqrt\gamma\,[(D\phi)^2]^2$ would otherwise shift the quartic coefficient. A useful example within the two-derivative class is
\begin{equation}
I_{\rm fin}=\alpha\int d^2x\sqrt{\gamma}\,D_a\phi D^a\phi\,.
\end{equation}
This term is Weyl invariant in two dimensions and preserves the constant shift symmetry. Substitution into the response formulas gives
\begin{align}
\Delta\langle T_{ab}\rangle
&=2\alpha\left[D_a\phi D_b\phi
-\frac12\gamma_{ab}D_c\phi D^c\phi\right],\\
\Delta\langle\mathcal O_\phi\rangle
&=-2\alpha D^2\phi\,.
\end{align}
These local shifts preserve the diffeomorphism Ward identity and the trace anomaly. At fixed boundary metric, the counterterm changes only a local quadratic term in the scalar generating functional; its nonlocal quadratic part and the terms of quartic and higher order are unaffected by this scheme change. The bulk asymptotic data $T_{ab}$ and $\psi$ themselves remain unchanged. Throughout the following analysis we use the reference scheme $\alpha=0$.

\section{Charged black brane: thermodynamics and holography}
\label{sec:gravity-prediction}\label{sec:charged-brane}
We now apply the holographic dictionary developed above to a charged AdS$_{3}$ black brane solution. First, we review the solution in the canonical ensemble using the scalar representation. 
Then, we determine its thermodynamic quantities from the renormalised stress tensor and show that the Euclidean on-shell renormalised action satisfies the quantum statistical relation
\begin{equation}
    I_{\rm{ren}}=\beta\mathcal F = \beta (M-TS)\ ,
\end{equation}
where $\mathcal F$ is the Helmholtz free energy, $T$ is the Hawking temperature and $S$ is the Bekenstein--Hawking entropy. The scalar $T=\gamma^{ab}T_{ab}$ of the preceding section will be written explicitly as $T^a{}_a$ here to distinguish it from temperature. We then identify the semiclassical scaling in which the Maxwell field produces finite backreaction at large central charge and show that the nonlinear charge dependence of the action is governed by Catalan numbers.

\subsection{Charged black brane action}

The Euclidean charged black brane metric \cite{Clement:1993kc, Martinez:1999qi} can be written as 
\begin{equation}\label{eq:brane-line}
    ds^{2}=f(r)d\tau^{2}+\frac{dr^{2}}{f(r)} +\frac{r^2}{\ell^2}dx^{2}\ ,
\end{equation}
The spatial coordinate $x$ is noncompact. We first evaluate homogeneous densities and use a fiducial length $2W$ to display extensive quantities. Thus $\int dx$ is represented by $2W$, without introducing physical end faces at $x=\pm W$. The limit $W\to\infty$ is understood at the level of densities, in the infrared prescription of \autoref{sec:Vari}. A generating functional for arbitrary variations of the unbounded source requires the corresponding infrared boundary terms to be controlled separately. The metric function is
\begin{equation}\label{eq:brane-fr}
f(r)
=
\frac{r^2-r_h^2}{\ell^2}
-
e^2\kappa^2\ell^2q^2
\log\frac{r}{r_h}\equiv \frac{r^2-r_h^2}{\ell^2}
-
a^2\log\frac{r}{r_h}\,.
\end{equation}
Here $r_{h}$ is the black brane horizon and $q$ is the electric charge density. The parameter $a=e\kappa\ell q$ is dimensionless. In Maxwell variables, the corresponding gauge field is
\begin{equation}\label{eq:brane-gaugefield}
    A_\tau(r)=-i e^2\ell q\log\frac{r}{r_h} \,,\quad F_{r\tau}=-\frac{i e^2\ell q}{r}\,.
\end{equation}
The condition $A_{\tau}(r_{h})=0$ ensures that the one-form is regular at the Euclidean horizon. 
Using the duality relation, the solution is described by 
\begin{equation}\label{eq:brane-scalar}
    \Phi(r,x)=\phi(x)=-e^{2}q x\,.
\end{equation}
One may alternatively identify $x\sim x+2W$. The Maxwell solution is periodic, while its scalar dual obeys the shift-twisted condition $\Phi(x+2W)=\Phi(x)-2We^2q$, so that $d\Phi$ is globally defined. This quotient leaves the local equations, Weyl anomaly, and classical thermodynamic densities of this saddle unchanged.
The non-compact correlator discussion is recovered in the decompactification limit. 

One can check that the equations of motion \eqref{eq:eom-Einstein-Scalar} and \eqref{eq:scalar} are satisfied component by component for the metric. 
The normalisation of the Maxwell field is consistent with the canonical momentum density. On a constant-$r$ surface, $n^r=\sqrt{f}$ and $\sqrt{\mathrm g}=r\sqrt{f}/\ell$, so
\begin{equation}
\Pi^\tau
=
\frac{1}{e^2}
\sqrt{\mathrm g}\,
n_\mu F^{\mu\tau}
=
-iq.
\end{equation}
Thus, fixing $q$ is equivalent to imposing the canonical boundary condition
\begin{equation}
    \delta\Pi^\tau=0.
\end{equation}
Expanding the metric near the horizon  $r=r_{h}$ and requiring regularity yields the Euclidean time period $\beta$ and the Hawking temperature $T$,
\begin{equation}\label{eq:brane-temperature}
T=\frac{1}{\beta}=\frac{r_h}{2\pi\ell^2}-\frac{a^2}{4\pi r_h}\ ,\quad 2r_h^2>a^2\ell^2\ .
\end{equation}
We consider the nonextremal saddle; equality in the last condition is reached only as the zero-temperature limit. With $\kappa^2=8\pi G$, the Bekenstein--Hawking entropy is
\begin{equation}
    S=\frac{\rm{Area}}{4 G}=\frac{4\pi W r_h}{\kappa^2\ell}\ .
\end{equation}
To compute the mass, we transform the solution to the FG coordinates \eqref{eq:FG} by using the following relation
\begin{equation}
    \frac{dr}{\sqrt{f(r)}}=-\frac{\ell\,dz}{z}\,.
\end{equation}
Asymptotically, this gives
\begin{equation}
    r(z)
={}
\frac{\ell^{2}}{z}
+
z
\left[
-\frac{a^{2}}{8}\log \frac{z^2}{\ell^{2}}
+
\frac{r_h^2}{4\ell^{2}}
-
\frac{a^{2}}{4}\log\frac{r_{h}}{\ell}
+
\frac{a^{2}}{8}
\right]
+
O\!\left(
\frac{z^3}{\ell^2}\log^2\frac{z}{\ell}
\right)\,.
\end{equation}
Substituting this into the metric gives the boundary metric $\gamma_{ab}=\delta_{ab}$ and
\begin{align}
\mathrm g_{\tau\tau}
&=\frac{\ell^2}{z^2}-\frac{r_h^2}{2\ell^2}
+\frac{a^2}{2}\log\frac{r_h}{\ell}+\frac{a^2}{4}
+\frac{a^2}{4}\log\frac{z^2}{\ell^2}
+O\!\left(\frac{z^2}{\ell^2}\log^2\frac{z}{\ell}\right),\\
\mathrm g_{xx}
&=\frac{\ell^2}{z^2}+\frac{r_h^2}{2\ell^2}
-\frac{a^2}{2}\log\frac{r_h}{\ell}+\frac{a^2}{4}
-\frac{a^2}{4}\log\frac{z^2}{\ell^2}
+O\!\left(\frac{z^2}{\ell^2}\log^2\frac{z}{\ell}\right).
\end{align}
The scalar has no radial dependence, so comparison with \eqref{eq:asymptotic-phi-new} yields
\begin{equation*}
\phi=-e^2qx,\qquad D^2\phi=0,\qquad\psi=0.
\end{equation*}
Thus the local scalar one-point function vanishes on the open cylinder. For a variation of the source slope, the regulated response has delta-function support at the spatial endpoints $x=\pm W$, as shown in \autoref{app:charge-variation}.
In the reference scheme $I_{\rm fin}=0$, matching the metric to \eqref{eq:asymptotic-metric-new} gives
\begin{align}\label{eq:brane-stresstensor}
\left\langle T_{\tau\tau}\right\rangle
=&
-\frac{r_h^2}{2\kappa^2\ell^3}
+
\frac{e^2\ell q^2}{2}
\log\frac{r_h}{\ell}\,,\\
\left\langle T_{xx}\right\rangle
=&
\frac{r_h^2}{2\kappa^2\ell^3}
-
\frac{e^2\ell q^2}{2}
\left(
1+\log\frac{r_h}{\ell}
\right) \ .
\end{align}
The off-diagonal component vanishes. These responses satisfy the constraints of \autoref{sec:AsymExp}:
\begin{equation*}
T^a{}_a=-\frac{e^2\ell q^2}{2}
=-\frac{\ell}{2e^2}D_a\phi D^a\phi,
\qquad D^bT_{ab}=0=-\psi D_a\phi.
\end{equation*}
The Euclidean time component has the opposite sign to the Lorentzian energy density. We therefore identify the extensive mass $M$ as
\begin{equation}\label{eq:mass}
    M = -\int_{-W}^{W} dx T^{\tau\tau}=2W\left(\frac{r_h^2}{2\kappa^2\ell^3}-\frac{e^2\ell q^2}{2}\log\frac{r_h}{\ell}\right)\ .
\end{equation}
For $q=0$, we recover the mass of the standard BTZ black brane. The logarithmic contribution is a characteristic feature of Maxwell theory in three dimensions. Since the electric field falls off as $1/r$, its radial energy integral is logarithmic. Changing the reference scale shifts the energy by a local term proportional to $q^{2}$, as expected from the finite counterterm discussed in \autoref{sec:einstein-maxwell}.

We now evaluate the canonical renormalised action directly.  Imposing the holographic cutoff at $z=\varepsilon$, corresponding to the large-radius surface $r= R_{\varepsilon}$, the on-shell value of the Einstein--scalar bulk action is
\begin{equation}\label{eq:onshell-bulk}
    I_{\rm bulk}=2 W\beta \frac{R_{\varepsilon}^{2}-r^2_{h}}{\kappa^{2}\ell^3}\,,
\end{equation}
whereas the Gibbons--Hawking term is
\begin{equation}\label{eq:onshell-gh}
    I_{\rm{GH}}=-\frac{2W \beta}{\kappa^{2}}\left[\frac{2 R^{2}_{\varepsilon}}{\ell^3}-\frac{a^2}{\ell}\log (R_{\varepsilon}/\ell) -   \frac{r_{h}^{2}}{\ell^3}+\frac{a^2}{\ell}\log (r_{h}/\ell) -\frac{a^2}{2\ell} \right]\ .
\end{equation}
The counterterm action, up to terms vanishing as $R_{\varepsilon}\to\infty$, is 
\begin{equation}\label{eq:onshell-counterterm}
    I_{ct}=\frac{2 W\beta }{\kappa^{2}}\left[\frac{R^{2}_{\varepsilon}}{\ell^3}-\frac{a^2}{\ell} \log (R_{\varepsilon}/\ell) -   \frac{r_{h}^{2}-a^2\ell^2\log (r_{h}/\ell)}{2\ell^3}\right]\,.
\end{equation}
Taking the cutoff limit of the sum of \eqref{eq:onshell-bulk}, \eqref{eq:onshell-gh}, and \eqref{eq:onshell-counterterm} gives
\begin{equation}\label{eq:canonical-saddle-action}
I_{\rm ren}=2\beta W\left[
-\frac{r_h^2}{2\kappa^2\ell^3}
+\frac{e^2\ell q^2}{2}
-\frac{e^2\ell q^2}{2}\log\frac{r_h}{\ell}\right].
\end{equation}
This satisfies the canonical quantum statistical relation
\begin{equation}
I_{\rm ren}=\beta M-S=\beta\mathcal F.
\end{equation}

\subsection{Semi-classical regime}
The limit relevant for finite classical backreaction should be distinguished from the validity of a weakly charged saddle. At fixed $\ell$ and $e$, the gravitational action scales as
\begin{equation}
I_{\rm EH}\sim O(\ell/\kappa^2)\sim O(c),
\qquad c=12\pi\ell/\kappa^2.
\end{equation}
Since $\phi=-e^2qx$, the matter contribution scales as $O(q^2)$ at fixed temperature and volume. Keeping the dimensionless backreaction parameter $a=e\kappa\ell q$ finite therefore selects
\begin{equation}\label{eq: limit}
\kappa\to0,\qquad\kappa^2q^2\ \text{fixed},
\qquad \ell,e,\beta\ \text{fixed}.
\end{equation}
In this limit the matter and gravitational saddle actions are both $O(c)$. At fixed $q=O(c^0)$, by contrast, the charge-dependent matter action is $O(1)$ and its backreaction tends to zero. This does not invalidate the classical weakly charged solution or its holographic description. It means that a complete result at absolute order $c^0$ would also require quantum corrections of that order; retaining finite backreaction is the scaling chosen for the present classical comparison.

We can now make the structure of the charged black brane action more explicit. Since the thermodynamic variables are the inverse temperature $\beta$ and the charge density $q$, the horizon equation can be solved exactly as 
\begin{equation}
    r_h=\frac{\pi\ell^2}{\beta}\left(1+\sqrt{1+\frac{1}{2}\chi^2}\right)\ ,\quad a=e\ell\kappa q\ ,\quad \chi=\frac{a\beta}{\pi\ell}\ .
\end{equation}
The square-root structure implies that the non-linear charge dependence of the renormalised action is governed by Catalan numbers. Defining
\begin{equation}
    C_{n} = \frac{1}{n+1}\binom{2n}{n}\,,
\end{equation}
the renormalised Euclidean action has the convergent small-charge expansion for $|\chi|<\sqrt{2}$,
\begin{equation}\label{eq:action-Catalan}
I_{\rm ren}
= -\frac{4\pi^2W\ell}{\beta\kappa^2} \left[ 1
+ \frac{\chi^2}{4} \log\left(\frac{2\pi\ell}{\beta}\right)
+ \sum_{n=2}^{\infty} \frac{(-1)^n C_{n-1}} {(n-1)8^n} \chi^{2n}\right].
\end{equation}
The leading term in \eqref{eq:action-Catalan} corresponds to the action of the BTZ black brane. The subleading terms are expansions in the dimensionless parameter $\chi$, which scales as
\begin{align}
    \chi\sim \kappa q\sim O(c^0)\ ,\quad c\to \infty\ .
\end{align}
At fixed $\chi$, each term in the expansion \eqref{eq:action-Catalan} is of the order $O(c)$. As we will discuss in the next section, the physical understanding of the result \eqref{eq:action-Catalan} is that the action of the charged black brane can be viewed as the sum of all integrated tree-level Witten diagrams against the source in the BTZ black brane background. Higher orders can involve graviton exchanges and graviton self-interactions. In the usual saddle expansion at fixed rescaled source, one-loop contributions are of absolute order $c^0$, or relative order $1/c$ compared with the classical action. 
They are not included here.

\section{Perturbative reconstruction from bulk correlation functions}
\label{sec:dual-CFT}
In this section, we reconstruct the charge expansion through quartic order using tree-level Witten diagrams around the neutral background. We organise the calculation as an expansion of the boundary generating functional in connected correlators. We use the sources $\gamma_{ab}$ and $\phi$, the fixed-flux prescription of \autoref{sec:Vari}, and the reference scheme $I_{\rm fin}=0$. The required dynamical data are supplied by the bulk theory: at quadratic order by the scalar propagator, and at quartic order by the complete graviton-exchange diagram. This calculation provides a consistency check of the holographic description. Matching these results to a concrete CFT remains a direction for future work, as we discuss in \autoref{sec:discussion}.

The quantum statistical relation established in the previous section is consistent with the semiclassical generating-functional prescription, with the same source sector and infrared treatment on both sides. For the static source $\phi=-e^2qx$, the thermal trace and the Euclidean generating functional are related by
\begin{equation}\label{eq: Ensemble=Generating}
    \operatorname{Tr}\!\left(e^{-\beta H_{\phi}}\right)
    =Z[\gamma,\phi]\,,
    \qquad \log Z[\gamma,\phi]
    =-I_{\rm ren}+O(c^0)\,,
\end{equation}
where $H_{\phi}$ is the Hamiltonian in the presence of the spatially varying marginal source. The generating functional can be written as
\begin{equation}
    Z[\gamma,\phi]
    =Z[\gamma,0]
    \left\langle
    e^{-\int d^2x\sqrt{\gamma}\phi\mathcal O_\phi}
    \right\rangle_{\rm cyl}\,.
\end{equation}
Here $\mathcal O_\phi$ is the operator introduced in \autoref{subsec:holo-ren-onepoint}, and the expectation value is evaluated in the undeformed theory on the thermal cylinder. Equivalently, after exchanging the spatial and Euclidean-time channels, it is the cylinder vacuum amplitude with the corresponding conditions at its two ends. The neutral saddle has a vanishing scalar field, but the scalar operator remains part of the boundary theory. Its central charge is $c=12\pi\ell/\kappa^2$ \cite{Brown:1986nw}.

It is convenient to compute the correlation functions on the complex plane. We use the conformal map
\begin{equation}\label{eq: mapping}
    \zeta=e^{2\pi w/\beta}\,,\qquad w=x+i\tau\,,
\end{equation}
under which the cylinder metric takes the form
\begin{equation}\label{eq: Weyl}
    ds^2=dwd\bar w=e^{-\sigma}d\widetilde s^2\,,
    \qquad d\widetilde s^2=\ell^2d\zeta d\bar\zeta\,.
\end{equation}
We have taken $\zeta$ to be dimensionless and retained $\ell$ as the reference length on the plane. Under a finite Weyl transformation, the partition function satisfies
\begin{equation}\label{eq:finite-weyl-Z}
    Z[e^{-\sigma}\widetilde\gamma,\phi]
    =e^{-\mathcal W[\sigma,\phi;\widetilde\gamma]}
    Z[\widetilde\gamma,\phi]\,.
\end{equation}
The induced action $\mathcal W$ is fixed by the Weyl anomaly. We first derive its matter contribution by conformal perturbation theory, using the two-point normalisation and source-shift identity supplied by the bulk scalar theory. We then evaluate the quadratic term and turn to the connected four-point function. To distinguish the generating functional from the induced action, we write $\mathcal G=\log Z$ throughout this section.

The expansion contains only even powers of the source. In addition to constant shifts, the Einstein--scalar action is invariant under $\Phi\to-\Phi$, and the neutral saddle and the counterterms used here preserve this symmetry. Hence $Z[\gamma,\phi]=Z[\gamma,-\phi]$ in this source sector, so all odd connected correlators evaluated at $\phi=0$ vanish, including their contact terms in a symmetry-preserving prescription. Diagrammatically, every vertex contains an even number of scalar legs. This conclusion follows from the reflection symmetry and the choice of state.

\subsection{Weyl anomaly and induced action from the marginal deformation}
\label{sec:weyl-induced}
The holographic trace identity in our conventions is
\begin{equation}\label{eq: trace anomaly}
    \left\langle T^a{}_a\right\rangle
    =\frac{c}{24\pi}\mathcal R[\gamma]
    -\frac{\ell}{2e^2}(\partial\phi)^2\,.
\end{equation}
The curvature term is the usual Weyl anomaly of the undeformed CFT. The source-dependent term can be understood from the short-distance singularities of conformal perturbation theory \cite{Jensen:2010em,Friedan:2012hi,Gomis:2015yaa}. To see this, consider an exactly marginal deformation with a local source $\lambda$. Constant $\lambda$ preserves conformal invariance, whereas a position-dependent source produces a local Weyl response. In the scalar sector of interest, this response can be written as
\begin{equation}
    \delta_\sigma\mathcal G[\widetilde\gamma,\lambda]
    -\delta_\sigma\mathcal G[\widetilde\gamma,0]
    =\int d^2x\sqrt{\widetilde\gamma}\,
    \delta\sigma\,g(\lambda)(\partial\lambda)^2\,.
\end{equation}
Here the Weyl variation is defined by $\delta\widetilde\gamma_{ab}=-\delta\sigma\widetilde\gamma_{ab}$. We use a covariant source scheme in which removable total-derivative terms have been fixed, and no additional vector sources are turned on. The function $g(\lambda)$ is proportional to the metric along the marginal direction. In a generic CFT it need not be constant.

In our case, the classical bulk action is invariant under a constant scalar shift. In the semiclassical source sector, and with compatible boundary terms and infrared data, the holographic generating functional inherits the identity
\begin{equation}\label{eq:source-shift-4}
    Z[\widetilde\gamma,\phi+\lambda_0]
    =Z[\widetilde\gamma,\phi]\,,
\end{equation}
where $\lambda_0$ is constant. The identity includes the contact terms required to define integrated correlators. It is a property inherited from the chosen bulk model, rather than an assumption about a general marginal operator. It implies that $g(\lambda)$ is constant along this direction. Consequently, the matter Weyl anomaly is quadratic in the source and is determined entirely by the two-point function. 

Conformal invariance fixes the separated-point dependence of the two-point function. Its coefficient follows from the bulk massless-scalar propagator and the coupling $e$:
\begin{equation}\label{eq:two-point-normalisation}
    \left\langle\mathcal O_\phi(x)\mathcal O_\phi(y)\right\rangle
    =\frac{C_{\mathcal O_\phi}}{|x-y|^4}\,,
    \qquad C_{\mathcal O_\phi}=\frac{2\ell}{\pi e^2}\,.
\end{equation}

To see the normalisation explicitly, consider Poincar\'e AdS$_3$ with metric $ds^2=\ell^2(dz^2+dx^a dx_a)/z^2$. The regular solution of the linear scalar equation with boundary source $\phi$ is
\begin{equation}\label{eq:scalar-bulk-boundary}
\Phi(z,x)=\int d^2y\,K(z,x;y)\phi(y),\qquad
K(z,x;y)=\frac1\pi\frac{z^2}{(z^2+|x-y|^2)^2}.
\end{equation}
The coefficient is fixed by $\int d^2y\,K(z,x;y)=1$. At separated points, comparison with \eqref{eq:asymptotic-phi-new} gives
\begin{equation}
\frac{\delta\psi(x)}{\delta\phi(y)}
=-\frac{2\ell}{\pi e^2|x-y|^4},\qquad x\ne y.
\end{equation}
Since the source enters as $+\int\phi\mathcal O_\phi$ in the Euclidean action, $\delta\langle\mathcal O_\phi\rangle/\delta\phi=-\langle\mathcal O_\phi\mathcal O_\phi\rangle_c$. This yields \eqref{eq:two-point-normalisation}; the logarithmic and finite local terms fix its contact extension separately.
Here $x$ and $y$ are local Cartesian coordinates with the physical flat metric. Consider its contribution to $\mathcal G$,
\begin{equation}\label{eq:two-point-integral}
    \mathcal G^{(2)}=\frac12\int d^2x\,d^2y\,
    \phi(x)\phi(y)
    \left\langle\mathcal O_\phi(x)\mathcal O_\phi(y)\right\rangle\,.
\end{equation}
The factor $1/2$ is the cumulant factor, so $\mathcal G^{(2)}$ contributes to $\log Z$ with a positive sign. The corresponding quadratic term of the Euclidean action is $-\mathcal G^{(2)}$. Writing $x=y+\rho$ and expanding the source at $y$ gives
\begin{equation}
    \phi(y+\rho)=\phi(y)
    +\rho^a\partial_a\phi(y)
    +\frac12\rho^a\rho^b\partial_a\partial_b\phi(y)+\cdots\,.
\end{equation}
With a hard-distance cutoff $|\rho|>r_0$, the angular integration yields
\begin{equation}\label{eq:two-point-divergences}
    \mathcal G^{(2)}_{\rm reg}
    =\frac{\pi C_{\mathcal O_\phi}}{2r_0^2}\int d^2y\,\phi^2
    -\frac{\pi C_{\mathcal O_\phi}}4\log(r_0\mu_{\rm R})
    \int d^2y\,\phi\Box\phi
    +\text{finite}\,.
\end{equation}
The arbitrary scale $\mu_{\rm R}$ makes the logarithm dimensionless. Counterterms cancel the two displayed divergences. The complete kernel, including the finite local term, is fixed by the Fourier-space prescription \eqref{eq:quadratic-fourier-kernel}. Its derivation in \autoref{app:quadratic-prescription} makes the correspondence with $I_{\rm fin}=0$ explicit.

\paragraph{Conformal transformation.}
The logarithmic divergence determines the Weyl response. Under $\gamma'=e^{-\sigma}\widetilde\gamma$, a fixed physical cutoff corresponds locally to
\begin{equation}
    r_\sigma(y)=e^{\sigma(y)/2}r_0\,,
    \qquad \mu_\sigma(y)=e^{-\sigma(y)/2}\mu_{\rm R}\,.
\end{equation}

Indeed, a coordinate separation $|\rho|$ in a locally flat patch of $\widetilde\gamma$ has physical length $e^{-\sigma(y)/2}|\rho|$ in $\gamma'$. Keeping that length equal to $r_0$ therefore moves the coordinate cutoff to $r_\sigma(y)$. The second relation, $r_\sigma\mu_\sigma=r_0\mu_{\rm R}$, expresses the same subtraction in the reference coordinates. Comparing the logarithmic subtractions gives $\log(\mu_\sigma/\mu_{\rm R})=-\sigma/2$. The remaining finite part is specified by \eqref{eq:quadratic-fourier-kernel}.

The logarithmic subtraction therefore produces
\begin{equation}
    -\frac{\pi C_{\mathcal O_\phi}}8
    \int d^2y\,\sigma\phi\Box\phi\,.
\end{equation}
In a covariant extension, the local curvature contact term
$\frac{\pi C_{\mathcal O_\phi}}{16}\int\sqrt\gamma\,\mathcal R[\gamma]\phi^2$
contributes the corresponding $\frac{\pi C_{\mathcal O_\phi}}{16}\int\phi^2\Box\sigma$. These pieces combine into
\begin{align}\label{eq:quadratic-weyl-response}
    \Delta \mathcal G^{(2)}_{\rm ren}
    &=-\frac{\pi C_{\mathcal O_\phi}}8
    \int d^2x\sqrt{\widetilde\gamma}\,\sigma
    \left[\phi\widetilde\Box\phi
    -\frac12\widetilde\Box(\phi^2)\right]\nonumber\\
    &=\frac{\ell}{4e^2}
    \int d^2x\sqrt{\widetilde\gamma}\,\sigma
    (\widetilde\partial\phi)^2\,.
\end{align}
The curvature contact term combines with the regulated two-point functional to enforce \eqref{eq:source-shift-4}. The logarithmic source is evaluated with the annular prescription in \autoref{app:quadratic-prescription}.

Adding the curvature anomaly of the undeformed CFT, we obtain
\begin{align}\label{eq:induced-action-full}
    \mathcal W[\sigma,\phi;\widetilde\gamma]
    ={}&\frac{c}{48\pi}
    \int d^2\zeta\sqrt{\widetilde\gamma}
    \left[\sigma\mathcal R[\widetilde\gamma]
    -\frac12(\widetilde\partial\sigma)^2\right]\nonumber\\
    &-\frac{\ell}{4e^2}
    \int d^2\zeta\sqrt{\widetilde\gamma}\,
    \sigma(\widetilde\partial\phi)^2\,.
\end{align}
With the stress-tensor convention of \autoref{subsec:holo-ren-onepoint},
\begin{equation}
    \left\langle T^a{}_a\right\rangle_{\gamma'}
    =\frac{2}{\sqrt{\gamma'}}\frac{\delta\mathcal W}{\delta\sigma}
    =\frac{c}{24\pi}\mathcal R[\gamma']
    -\frac{\ell}{2e^2}(\partial\phi)^2_{\gamma'}\,,
\end{equation}
in agreement with \eqref{eq: trace anomaly}. In particular, there is no quartic matter contribution to this Weyl anomaly. For smooth sources, the quartic term has no local Weyl variation in this scheme.

\subsection{Quadratic order in the charge density}
\label{sec:quadratic-revision}
Under \eqref{eq: mapping}, the regulated cylinder $-W<x<W$ is mapped to the annulus
\begin{equation}
    \eta<|\zeta|<\eta^{-1}\,,\qquad
    \mathcal L\equiv\log\frac1\eta=\frac{2\pi W}{\beta}\,.
\end{equation}
The parameter $\mathcal L$ is the dimensionless cylinder half-length, and $\eta$ is an infrared regulator, distinct from the bulk radial cutoff $\varepsilon$. The Weyl factor and source on the plane are
\begin{equation}\label{eq:annulus-source}
    \sigma=2\log\left(\frac{2\pi\ell}{\beta}|\zeta|\right)\,,
    \qquad
    \phi=-\frac{e^2\beta q}{2\pi}\log|\zeta|\,.
\end{equation}
For later use, we introduce the unit-normalised operator and its conjugate source,
\begin{equation}\label{eq:normalised-source}
    \widehat{\mathcal O}_\phi=\frac{\mathcal O_\phi}{\sqrt{C_{\mathcal O_\phi}}}\,,
    \qquad J=\sqrt{C_{\mathcal O_\phi}}\phi=p\log|\zeta|\,,
    \qquad p^2=\frac{e^2\ell\beta^2q^2}{2\pi^3}\,.
\end{equation}
The sign of $p$ follows from \eqref{eq:annulus-source} and will not affect the even-order terms. When using dimensionless plane coordinates, the conformal factor $\ell^2$ is included in the operator insertion, so that the correlator in coordinate measure is $\langle\widehat{\mathcal O}_\phi(\zeta)\widehat{\mathcal O}_\phi(0)\rangle=|\zeta|^{-4}$. Thus no factors of $\ell$ are hidden in the source integrals below.


The source is dimensionless and has zero Weyl weight, so no additional factor accompanies it under the map. Write $b=e^2\beta q/(2\pi)$, so that $\phi=-b\log|\zeta|$ and $p^2=C_{\mathcal O_\phi}b^2$. With $\rho=|\zeta|$ and $d\widetilde s^2=\ell^2(d\rho^2+\rho^2d\theta^2)$,
\begin{equation}
(\widetilde\partial\phi)^2=\frac{b^2}{\ell^2\rho^2},
\qquad
\sqrt{\widetilde\gamma}\,d^2\zeta=\ell^2\rho\,d\rho\,d\theta.
\end{equation}
Thus the factors of $\ell$ in the measure and inverse metric cancel. The quadratic Weyl response reduces to
\begin{equation}
\mathcal W_{\rm matter}
=-\frac{\ell b^2}{4e^2}
\int_0^{2\pi}d\theta\int_{e^{-\mathcal L}}^{e^{\mathcal L}}\frac{d\rho}{\rho}
\,2\left[\log\frac{2\pi\ell}{\beta}+\log\rho\right].
\end{equation}
This makes explicit which part of the quadratic result follows from the two-point anomaly.

The relevant integrals on the symmetric annulus are
\begin{equation}
    \int_{\eta<|\zeta|<\eta^{-1}}\frac{d^2\zeta}{|\zeta|^2}
    =4\pi\mathcal L\,,\qquad
    \int_{\eta<|\zeta|<\eta^{-1}}
    \frac{d^2\zeta}{|\zeta|^2}\log|\zeta|=0\,.
\end{equation}
We apply the Weyl transformation to $\mathcal G[\gamma,\phi]-\mathcal G[\gamma,0]$, so the neutral end contributions cancel. The matter term in \eqref{eq:induced-action-full} gives
\begin{equation}\label{eq:induced-on-annulus}
    \mathcal W_{\rm matter}
    =-\pi^2p^2\mathcal L\log\left(\frac{2\pi\ell}{\beta}\right).
\end{equation}
We fix the neutral normalisation separately by propagating the cylinder vacuum of energy $-c/12$ over the length $2\mathcal L$. This gives $\mathcal G^{(0)}=c\mathcal L/6+o(\mathcal L)$.

The finite plane pairing can be evaluated with the same subtraction. In dimensionless Poincar\'e coordinates, the regular scalar profile
\begin{equation}\label{eq:log-source-extension}
\Phi(u,\rho)=-\frac b2\log(u^2+\rho^2)
\end{equation}
approaches $\phi=-b\log\rho$. We retain its induced data at the inner and outer annular surfaces and remove the bulk radial cutoff before taking $\mathcal L\to\infty$. Including both end contributions, \autoref{app:quadratic-prescription} gives
\begin{equation}\label{eq:plane-pairing-zero}
I_{\rm plane,ren}^{(2)}
=\frac{\pi\ell b^2}{2e^2}
\left[(\log\rho_+)^2-(\log\rho_-)^2\right]=0,
\qquad \rho_\pm=e^{\pm\mathcal L},\quad \alpha=0.
\end{equation}
This specifies the annular extension of the renormalised two-point pairing. Combining it with the matter Weyl response and the neutral vacuum term gives
\begin{equation}\label{eq:quadratic-match-revision}
    \mathcal G^{(0)}+\mathcal G^{(2)}
    =\left[\frac c6+
    \pi^2p^2\log\left(\frac{2\pi\ell}{\beta}\right)\right]\mathcal L
    +o(\mathcal L)\,.
\end{equation}
In the original variables this gives
\begin{equation}
    I_{\rm ren}^{(0)}+I_{\rm ren}^{(2)}
    =-\frac{4\pi^2W\ell}{\beta\kappa^2}
    -e^2\ell\beta Wq^2\log\left(\frac{2\pi\ell}{\beta}\right)\,,
\end{equation}
which reproduces \eqref{eq:action-Catalan}. Adding the finite counterterm discussed in \autoref{subsec:holo-ren-onepoint} would shift this action by
\begin{equation}
    \Delta I_{\rm ren}^{(2)}=2\alpha\beta We^4q^2\,.
\end{equation}
The same finite term therefore shifts the quadratic contribution in both descriptions. Once the reference scheme is fixed, there is no further freedom of this form to change the quartic coefficient.

\subsection{Quartic order in the charge density}
\label{sec:quartic-revision}
The quartic term pairs the connected four-point function with four copies of the logarithmic source. A conformal partial-wave representation factorises the source integrations into two pair integrals. We first fix the operator and partial-wave conventions, then introduce the graviton-exchange kernel and evaluate its analytic source pairing.

With the normalisation in \eqref{eq:normalised-source}, define
\begin{equation}\label{eq:quartic-kernel-definition}
    \left\langle\widehat{\mathcal O}_\phi(\zeta_1)\widehat{\mathcal O}_\phi(\zeta_2)
    \widehat{\mathcal O}_\phi(\zeta_3)\widehat{\mathcal O}_\phi(\zeta_4)\right\rangle_c
    =\frac1c\mathcal H(\zeta_i)+O(c^{-2})\,.
\end{equation}
The fourth-order contribution to the generating functional is
\begin{equation}\label{eq:quartic-cumulant}
    \mathcal G^{(4)}_{\rm AC}
    =\frac{p^4}{4!c}\,
    \operatorname{AC}_{\mathcal L}\!\int\prod_{i=1}^{4}d^2\zeta_i\,
    \mathcal H(\zeta_i)\prod_{i=1}^{4}\log|\zeta_i|\,.
\end{equation}
The subscript $c$ in \eqref{eq:quartic-kernel-definition} removes disconnected products of two-point functions. The positive sign and the factor $1/4!$ in \eqref{eq:quartic-cumulant} follow from the connected expansion of $\log Z$. The notation $\operatorname{AC}_{\mathcal L}$ specifies the analytic source pairing and infrared prescription defined below; the unregulated integral is not absolutely convergent.

\paragraph{Conformal partial waves and shadow normalisation.}

The partial-wave representation separates the conformal kinematics from the dynamical coefficients $b_j(\nu)$ \cite{Dolan:2003hv,Dolan:2011dv,Costa:2012cb}. In two Euclidean dimensions the principal series is labelled by a dimension $\Delta=1+i\nu$ and integer spin $j$. A conformal partial wave combines the global block of this representation with that of its shadow, of dimension $2-\Delta=1-i\nu$. These contributions have the same quadratic Casimir eigenvalue; their relative coefficient is fixed by the single-valued Euclidean integral. The principal-series integral supplies a basis for the Euclidean four-point function.

The shadow representation expresses each partial wave as an integral of two three-point structures over a common boundary point $y$ \cite{Osborn:2012vt}. At $y$, the two structures carry weights $(h,\bar h)$ and $(1-h,1-\bar h)$, so their product has weights $(1,1)$ and can be integrated with the measure $d^2y$. At fixed $y$, the four source integrations factorise into two independent integrals, one for each pair of insertions. We normalise the resulting waves to match the convention used in the graviton-exchange formula \cite{Costa:2014kfa}.
To perform the source integrals, we express the partial waves as boundary shadow integrals. let us introduce
\begin{equation}\label{eq:shadow-weights}
    h_j=\frac{1+i\nu+j}{2}\,,\qquad
    \bar h_j=\frac{1+i\nu-j}{2}\,,\qquad j=0,\pm2\,.
\end{equation}
The bar on $\bar h_j$ labels the antiholomorphic weight; it does not denote complex conjugation. For external weights $(1,1)$, the three-point structure without an OPE coefficient is
\begin{equation}\label{eq:shadow-three-point-revision}
    \mathcal T_{h,\bar h}(12;y)
    =\frac{\zeta_{12}^{h-2}\bar \zeta_{12}^{\bar h-2}}
    {(\zeta_1-y)^h(\zeta_2-y)^h
    (\bar \zeta_1-\bar y)^{\bar h}(\bar \zeta_2-\bar y)^{\bar h}}\,.
\end{equation}
The raw shadow integral is
\begin{equation}\label{eq:raw-shadow}
    \Psi^{12|34}_{\nu,j}
    =\operatorname{AC}\int_{\mathbb C}d^2y\,
    \mathcal T_{h_j,\bar h_j}(12;y)
    \mathcal T_{1-h_j,1-\bar h_j}(34;y)\,.
\end{equation}
Here $d^2y=d(\operatorname{Re}y)d(\operatorname{Im}y)$. The second factor carries the shadow dimension and opposite helicity. We define the normalised waves used in the exchange representation by
\begin{equation}\label{eq:shadow-wave-conversion}
    \mathcal F^{12|34}_{\nu,0}=\frac{\Psi^{12|34}_{\nu,0}}{B_0(\nu)}\,,
    \qquad
    \mathcal F^{12|34}_{\nu,2}
    =\frac{\Psi^{12|34}_{\nu,2}+\Psi^{12|34}_{\nu,-2}}{B_2(\nu)}\,,
\end{equation}
with
\begin{align}\label{eq:shadow-normalisations-revision}
    B_0(\nu)&=\frac{128\cosh^4(\pi\nu/2)}
    {\pi\nu\sinh(\pi\nu)(\nu^2+1)^2}\,,\nonumber\\
    B_2(\nu)&=\frac{8192\nu\cosh^4(\pi\nu/2)}
    {\pi\sinh(\pi\nu)(\nu^2+1)^3(\nu^2+9)^2}\,.
\end{align}
These factors follow by matching the physical and shadow OPE branches of \eqref{eq:raw-shadow} to the block convention \eqref{eq:quartic-global-blocks}. The explicit ratios and their gamma-function reduction are given in \autoref{app:quartic-details}.

\paragraph{The bulk exchange kernel.}
The required partial-wave decomposition is obtained by specialising the graviton-exchange Witten diagram of \cite{Costa:2014kfa} to boundary dimension $d=2$ and external scalar dimension $\Delta_{\mathcal O_\phi}=2$. Removing the overall factor $1/c$, its direct-channel contribution is
\begin{equation}\label{eq:exchange-partial-waves}
    \mathcal H_{12|34}
    =\operatorname{AC}\int_{-\infty}^{\infty}d\nu\,
    \left[b_2(\nu)\mathcal F^{12|34}_{\nu,2}
    +b_0(\nu)\mathcal F^{12|34}_{\nu,0}\right],
\end{equation}
where $\mathcal F_{\nu,j}$ are the waves defined in \eqref{eq:shadow-wave-conversion}. With unit-normalised external operators, the graviton-exchange coefficients are \cite{Costa:2014kfa}
\begin{equation}\label{eq:bulk-spectral-coefficients}
    b_2(\nu)=\frac6{\nu^2+1}\,,\qquad
    b_0(\nu)=-\frac3{16}\frac{(\nu^2+1)^2}{\nu^2+4}\,.
\end{equation}
The coupling conversion $\kappa^2/\ell=12\pi/c$ and the resulting coefficients are derived in \eqref{eq:exchange-coefficients-coupling}. The spectral measure is $d\nu$ on the full real axis, with no additional factor $1/(2\pi)$. The complete tree-level kernel is
\begin{equation}\label{eq:exchange-channel-sum}
    \mathcal H=\mathcal H_{12|34}+\mathcal H_{13|24}+\mathcal H_{14|23}\,.
\end{equation}
The sum runs over the three distinct pairings of the external scalar legs.

The spin-two and spin-zero terms arise from the transverse traceless and trace contributions to the graviton propagator, respectively \cite{Costa:2014kfa}. Both enter the exchange between conserved scalar stress tensors. The stress-tensor Ward identity checks the spin-two pole coefficient, as shown in \eqref{eq:stress-block-check}, while the complete diagram fixes the regular terms. 

\paragraph{Prescription for the source integral.}
We compute the action per unit spatial length on the infinite thermal cylinder. At fixed sewing point $y$, we integrate each pair of insertions over the full plane by analytic continuation, with the local prescription of \autoref{app:quartic-details} and no finite quartic contact term.

The product of the two vertices leaves the measure $d^2y/|y|^2$. Under the cylinder map, $s=\log|y|$ is the spatial coordinate in units of $\beta/(2\pi)$. For a fiducial length $2W$, the remaining integral is therefore
\begin{equation}\label{eq:common-scale-cutoff}
\int_{e^{-\mathcal L}<|y|<e^{\mathcal L}}\frac{d^2y}{|y|^2}
=2\pi\int_{-\mathcal L}^{\mathcal L} ds
=4\pi\mathcal L,\qquad \mathcal L=\frac{2\pi W}{\beta}.
\end{equation}
This cutoff on the common coordinate $y$, applied after the pair integrations, defines $\operatorname{AC}_{\mathcal L}$ in \eqref{eq:quartic-cumulant}.

\paragraph{Integration against the logarithmic source.}
For each pair of insertions, we define
\begin{equation}\label{eq:source-vertex-revision}
    V_j(\nu;y)=\operatorname{AC}\int d^2\zeta_1d^2\zeta_2\,
    \log|\zeta_1|\log|\zeta_2|\,\mathcal T_{h_j,\bar h_j}(12;y)\,.
\end{equation}
The complex beta identity \eqref{eq:complex-beta-revision} gives a vanishing integral when either source leg is constant, as evaluated in \eqref{eq:quartic-constant-leg}. Rescaling $\zeta_i=y u_i$ therefore yields
\begin{equation}
    V_j(\nu;y)=v_j(\nu)y^{-h_j}\bar y^{-\bar h_j}\,.
\end{equation}
At $y=1$, a change of variables reduces each logarithmic source to a difference of two plane Green functions. Applying the Laplacian to each source gives (\autoref{app:quartic-details})
\begin{equation}\label{eq:quartic-vertex-derivative}
    v_j(\nu)=-\frac{\pi^2}{2h_j\bar h_j(h_j-1)(\bar h_j-1)}\,.
\end{equation}
Substituting \eqref{eq:shadow-weights} yields
\begin{equation}\label{eq:source-vertices-result}
    v_0(\nu)=-\frac{8\pi^2}{(\nu^2+1)^2}\,,\qquad
    v_2(\nu)=v_{-2}(\nu)
    =-\frac{8\pi^2}{(\nu^2+1)(\nu^2+9)}\,.
\end{equation}
The shadow vertex has the same coefficient. Using the spatial volume factor in \eqref{eq:common-scale-cutoff}, a single exchange diagram contributes
\begin{equation}\label{eq:one-channel-smearing-revision}
    \operatorname{AC}_{\mathcal L}\!\int\prod_{i=1}^4d^2\zeta_i\,
    \mathcal H_{12|34}(\zeta_i)\prod_{i=1}^4\log|\zeta_i|
    =4\pi\mathcal L\int_{\mathbb R}d\nu\,R(\nu)\,,
\end{equation}
where
\begin{align}\label{eq:spectral-density-revision}
    R(\nu)
    &=\frac{2b_2(\nu)v_2(\nu)^2}{B_2(\nu)}
    +\frac{b_0(\nu)v_0(\nu)^2}{B_0(\nu)}\nonumber\\
    &=\frac{3\pi^5}{4}
    \frac{\tanh(\pi\nu/2)}
    {\nu(\nu^2+4)\cosh^2(\pi\nu/2)}\,.
\end{align}
The factor two counts the two spin-two helicities in \eqref{eq:shadow-wave-conversion}. The spin-zero completion is also essential to the second equality. The final integrand is regular at $\nu=0$ and decays exponentially, so no finite-part operation remains in the spectral integral.

For completeness, the remaining integration can be performed using
\begin{equation}
    \frac1{\nu(\nu^2+4)}
    =\frac14\lim_{\delta\to0^+}\int_0^\infty dX\,
    e^{-\delta X}(1-e^{-2X})\sin(\nu X)\,,
\end{equation}
and the Fourier transform
\begin{equation}
    \int_{\mathbb R}d\nu\,
    \frac{\cos(\nu X)}{\cosh^2(\pi\nu/2)}
    =\frac{4X}{\pi\sinh X}\,.
\end{equation}
The latter follows from Euler's beta integral after setting $t=e^{\pi\nu}$. An integration by parts gives
\begin{align}\label{eq:spectral-integral-revision}
    \int_{\mathbb R}d\nu\,
    \frac{\tanh(\pi\nu/2)}
    {\nu(\nu^2+4)\cosh^2(\pi\nu/2)}
    &=\frac1{\pi^2}\int_0^\infty dX\,
    \frac{X^2(1-e^{-2X})}{\sinh X}\nonumber\\
    &=\frac2{\pi^2}\int_0^\infty dX\,X^2e^{-X}
    =\frac4{\pi^2}\,.
\end{align}
The damping factor justifies the intermediate interchange; exponential decay and regularity at $\nu=0$ permit its removal. Hence $\int_{\mathbb R}R(\nu)d\nu=3\pi^3$. Each of the three exchange diagrams has the same pairing with four identical sources, and \eqref{eq:quartic-cumulant} becomes
\begin{equation}\label{eq:quartic-final-revision}
    \mathcal G^{(4)}_{\rm AC}
    =\frac{p^4}{24c}\,3\,(4\pi\mathcal L)\,(3\pi^3)
    =\frac{3\pi^4p^4}{2c}\mathcal L\,.
\end{equation}
Substituting \eqref{eq:normalised-source}, $c=12\pi\ell/\kappa^2$ and $\mathcal L=2\pi W/\beta$, the corresponding action contribution is
\begin{equation}\label{eq:quartic-action-revision}
    I^{(4)}_{\rm AC}=-\mathcal G^{(4)}_{\rm AC}
    =-\frac{e^4\kappa^2\ell\beta^3Wq^4}{16\pi^2}
    =-\frac{4\pi^2W\ell}{\beta\kappa^2}\frac{\chi^4}{64}\,.
\end{equation}
Dividing by $2W$ gives the same quartic action density as the expansion of the charged saddle in \eqref{eq:action-Catalan}.

\section{Discussions and outlook}\label{sec:discussion}

In this work, we have studied the charged AdS$_3$ black brane in the canonical ensemble. The scalar representation of the Maxwell field makes the boundary datum particularly simple: the electric charge density fixes the slope of a marginal scalar source. Holographic renormalisation then determines the one-point functions and the thermodynamic quantities, and the renormalised action satisfies the canonical quantum statistical relation. At fixed temperature, its charge expansion has nonlinear coefficients governed by Catalan numbers.

We have examined how the first terms of this expansion arise in perturbation theory around the neutral geometry. The two-point function, with its normalisation fixed by the bulk scalar action, determines the matter contribution to the Weyl anomaly. The regulated annular pairing in the reference scheme, evaluated in \autoref{app:quadratic-prescription}, then reproduces the quadratic term. At fourth order, the dynamical input is the complete graviton-exchange Witten diagram of the minimally coupled Einstein--scalar theory. With the common-scale cutoff \eqref{eq:common-scale-cutoff}, the integrated diagram reproduces the quartic term of the charged-brane action.

The comparison requires a common treatment of finite terms and infrared effects. We used the reference scheme $\alpha=0$ of \autoref{subsec:holo-ren-onepoint}. For the linear source, the finite counterterm $\alpha\int\sqrt{\gamma}(\partial\phi)^2$ shifts the Euclidean action by $2\alpha\beta We^4q^2$. It shifts the energy and the canonical free energy consistently, leaving the quantum statistical relation intact. This scheme dependence affects the quadratic coefficient, while the quartic coefficient is unchanged by this counterterm. Finite source functionals with more than two boundary derivatives are assumed zero in the prescription of \autoref{sec:Holography}.

The absence of odd powers follows from the reflection symmetry $\Phi\to-\Phi$ of the bulk action, the neutral saddle and our counterterm prescription. Consequently, odd connected scalar correlators vanish at zero source, including their contact terms in a symmetry-preserving scheme. The constant-shift symmetry has a different consequence: in the semiclassical source sector considered here, the metric along the marginal direction is independent of the scalar source. The matter Weyl anomaly is therefore quadratic and is completely determined by the two-point normalisation. At higher orders, the source-dependent contributions obey a homogeneous Weyl identity for smooth sources with no edge contributions. The corresponding connected correlators transform with the usual operator weights and retain non-trivial dependence on conformal cross-ratios.

These results suggest several directions for further investigation.

\paragraph{Asymptotic charges.}
An extension of the density calculation would be to determine the falloff conditions needed to define integrated asymptotic charges on the infinite line. Specifying these conditions would allow the thermodynamic analysis of densities to be extended to a phase space with well-defined global charges and an associated asymptotic symmetry algebra.

\paragraph{A concrete CFT description.}
Matching the charged-brane expansion to a concrete CFT remains an open problem. One approach is to first reconstruct the required correlators from boundary consistency conditions. Starting from a large-$c$ CFT with a dimension-two scalar operator and a specified single-trace spectrum, one would impose crossing symmetry and the stress-tensor Ward identities. At order $1/c$, the connected four-point function contains both stress-tensor exchange and corrections to the double-trace families $[\widehat{\mathcal O}_\phi\widehat{\mathcal O}_\phi]_{n,j}$, whose leading dimensions are $4+2n+j$. Determining their anomalous dimensions and OPE coefficients, together with the remaining contact ambiguity, would provide the data needed for an independent comparison.

A Mellin representation may help organise this reconstruction through its exchange poles, crossing properties and large-Mellin-variable behaviour \cite{Heemskerk:2009pn,Mack:2009mi,Costa:2012cb}. In two dimensions, the twist-zero stress tensor requires particular care with the continuation and contour prescription. A growth condition compatible with a two-derivative bulk theory, together with a contact-term prescription implementing the scalar source shift identity, could further constrain the answer. Establishing how strongly these conditions determine the correlator would help identify the boundary data that a concrete CFT realisation such as the one discussed in \cite{Jensen:2010em} must reproduce.

\paragraph{Higher-point functions and the Catalan expansion.}
The higher Catalan coefficients provide targets for extending the perturbative calculation. At sixth order, the charged solution predicts
\begin{equation}\label{eq:sixth-target}
    \mathcal G^{(6)}
    =-\frac{9\pi^6p^6}{2c^2}\mathcal L
\end{equation}
for the corresponding extensive contribution in a compatible prescription. Testing this coefficient requires the complete connected six-point function, including all tree diagrams allowed by the coupled Einstein--scalar action. More generally, it would be interesting to determine whether the Catalan pattern admits a direct explanation in terms of the integrated correlators, perhaps through a recursion relation that organises the higher-point contributions. Such a relation would connect the closed-form charged solution to the perturbative structure of the generating functional at arbitrary order, as well as relate tree-level Witten diagrams of fixed order amongst each other in possibly novel ways.

\paragraph{Other ensembles and symmetry resolution.}
Changing the Maxwell boundary conditions provides a natural extension of the present analysis. Fixing the renormalised boundary gauge field defines the grand-canonical problem, with the chemical potential encoded in the thermal holonomy. A scalar formulation of this ensemble must retain the global gauge data and allow the charge sector to vary with its chemical-potential weight. The linear profile $\phi=-e^2qx$ illustrates the issue: its radial derivative vanishes for every $q$, so homogeneous scalar Neumann data do not distinguish the different charge sectors. The relation between the ensembles, which has partially already been discussed in the Maxwell frame in \cite{Jensen:2010em},  therefore involves the winding and holonomy sectors of Abelian duality \cite{Beasley:2014vma,Beasley:2014kva}.

A treatment of these global sectors would also provide a starting point for charge-projected partition functions and symmetry-resolved entanglement. These applications require specifying the compactness of the gauge group, the allowed charges and holonomies, and their implementation on the relevant replica geometries. They would extend the present thermodynamic analysis to observables that resolve the charge dependence of quantum entanglement. Furthermore, charge-projected partition functions can, via a conformal transformation \cite{casini2011towards}, be mapped to the thermal partition function of a topological charged black hole. Applying our method to the charge expansion of the latter would potentially allow for a novel interpretation of the charge-projected partition functions in the CFT dual of Einstein-Maxwell in three dimensions in terms of connected correlation functions of the dual scalar operator.

\paragraph{Weyl transformations and logarithmic mixing.}
More general source profiles and Weyl-related boundary backgrounds would probe further information in the generating functional. The induced action in \autoref{sec:weyl-induced} determines the local anomalous response, while higher-order terms contain additional nonlocal information. In particular, the logarithmic backreaction of the Maxwell field modifies the transformation law of the holographic stress tensor. A boundary Weyl transformation mixes the finite metric coefficient with the coefficient of the radial logarithm. For a constant rescaling $\gamma'_{ab}=e^{2\omega}\gamma_{ab}$, with the scalar source unchanged, this gives
\begin{equation}
T'_{ab}
=T_{ab}-\frac{2\omega}{\kappa^2\ell}h_{(2)ab}
=T_{ab}+\frac{\ell\omega}{e^2}
\left(D_a\phi D_b\phi-\frac12\gamma_{ab}(D\phi)^2\right).
\end{equation}
For a local conformal map, the compensating Weyl parameter involves the logarithm of the conformal Jacobian. The stress-tensor transformation therefore contains a source-dependent logarithmic contribution in addition to the usual Schwarzian term. Its coefficient is fixed by the matter Weyl anomaly, and it vanishes for a constant scalar source.

This mixing resembles the triangular transformation law of a logarithmic pair, in which one field transforms homogeneously while its partner acquires an additive logarithm multiplying the first \cite{Gurarie:1993xq,Ghezelbash:1998rj,Grumiller:2010rm,Cardy:2013rqg}. Here $h_{(2)ab}$ is a local functional of the prescribed scalar source. Whether this resemblance extends to a logarithmic multiplet of boundary operators is an interesting question that requires further information about the spectrum and correlation functions.

\paragraph{Near-extremal regime.}
The exact charged-brane action also admits a near-extremal expansion. It would be interesting to understand how the expansion in integrated connected correlators studied here organises in this regime. Holographic Wilsonian renormalisation provides a concrete framework for addressing this question: integrating out the exterior geometry produces an effective action for fields on a radial matching surface near the entrance to the throat \cite{Faulkner:2010jy, Heemskerk:2010hk}. Related matching methods have been used to connect near-horizon correlators dressed by the Schwarzian mode to observables at the asymptotic boundary \cite{Daguerre:2023cyx}.

In our setting, one could follow the radial evolution of the source-dependent generating functional and ask how its connected-correlator expansion reorganises into the couplings and correlators of the near-horizon effective theory. Relating this description to the small-charge Catalan series would require a resummation, since that series is organised around the neutral saddle at fixed temperature. The full three-dimensional renormalised action would supply the UV matching data, including the local boundary terms discussed in \cite{Castro:2025itb}. This could clarify how the nonlinear charge dependence found here is encoded in the effective two-dimensional description of near-extremal thermodynamics. 

\paragraph{Integrability and non-compact boundaries.}
Black branes provide a natural setting for investigating gravitational dynamics with non-compact conformal boundaries, complementing the better-studied cases of global AdS$_3$ and BTZ geometries, whose conformal boundaries are compact. For pure three-dimensional gravity, such non-compact boundary geometries have recently been investigated from the perspective of integrable systems \cite{Adami:2026jmo,Adami:2025pfk}, with the associated phase spaces and boundary dynamics exhibiting structures related to integrable hierarchies and their evolution equations. The charged AdS$_3$ black branes studied here extend this setting by incorporating a $\mathrm{U}(1)$ gauge sector. Their interpretation in terms of marginally deformed two-dimensional conformal field theories provides a complementary perspective from which to investigate the interplay between gauge-field dynamics, boundary conditions, and integrability. A systematic extension of the integrable framework developed for pure gravity to the charged case would therefore be of particular interest. In particular, it would be worthwhile to determine whether the gauge sector admits a natural incorporation into the existing hierarchies, how the corresponding boundary evolution equations are modified by the presence of charge, and whether the marginal deformation has a direct interpretation in terms of the resulting integrable structures. Such an analysis could establish a more precise connection between the deformed-CFT description of charged black branes and the integrability of three-dimensional gravity with non-compact boundaries, while clarifying which features of the pure-gravity framework persist in the presence of additional matter degrees of freedom.

We hope to address some of these questions in the near future.

\acknowledgments
We thank Horacio Casini, Felipe Diaz, Hao Geng, Peng-Xiang Hao, Kristan Jensen, Bong Lian, Wenjie Ma, and Anna Tokareva for comments, suggestions, and useful conversations. The work of GAH was supported by the Beijing Natural Science Foundation (IS24016) and National Natural Science Foundation of China (NSFC) grant No. 12505091. SZ and YL acknowledge the support from the startup grant at Shanghai Institute for Mathematics and Interdisciplinary Sciences (SIMIS). RM acknowledges hospitality from SIMIS, and associated travel support under STCSM Grant 25HB2701900. RM furthermore acknowledges the support of the German Research Foundation (DFG) through the Collaborative Research Center ToCoTronics, Project-ID 258499086 (SFB 1170), as well as Germany's Excellence Strategy through the W{\"u}rzburg-Dresden Cluster of Excellence on Complexity, Dynamics and Topology in Quantum Matter - ctd.qmat (EXC 2147, Project-ID 390858490). We acknowledge the use of OpenAI's ChatGPT as a research tool, primarily in connection with the computation of the quartic contribution to the generating functional, as well as to help identify errors, work through intermediate steps in calculations and derivations, and suggest textual edits. All AI-assisted calculations, derivations, and numerical results were independently checked by the authors, who take full responsibility for the final manuscript.

\appendix

\section{Details of holographic renormalisation}\label{app:holo-ren}
We derive the radial coefficients, cancellation of divergences and finite responses used in \autoref{sec:einstein-maxwell}. We use the Fefferman--Graham gauge, outward normal and extrinsic curvature of \eqref{eq:FG}--\eqref{eq:induced-extrinsic-new}, and the reference scheme $I_{\rm fin}=0$. The general renormalisation procedure follows \cite{deHaro:2000xn,Skenderis:2002wp}; here we display the steps specific to the massless scalar and our conventions.

\subsection{Radial coefficients and constraints}
Before identifying the response functions, write the asymptotic fields as
\begin{align}\label{eq:app-expansions}
    \mathrm g_{ab}
    &=\frac{\ell^2}{z^2}\gamma_{ab}
    +\gamma_{(2)ab}+h_{(2)ab}L_z
    +O\!\left(\frac{z^2}{\ell^2}L_z^2\right),\nonumber\\
    \Phi
    &=\phi+\frac{z^2}{\ell^2}
    \left[\phi_{(2)}+\widetilde\phi_{(2)}L_z\right]
    +O\!\left(\frac{z^4}{\ell^4}L_z^2\right),
    \qquad L_z:=\log\frac{z^2}{\ell^2}\,.
\end{align}
All traces and boundary contractions below use $\gamma_{ab}$. Defining $\bar g_{ab}=z^2\mathrm g_{ab}/\ell^2$, the scalar equation becomes
\begin{equation}
    \partial_z\!\left(z^{-1}\sqrt{\bar g}\,\partial_z\Phi\right)
    +z^{-1}\partial_a\!\left(\sqrt{\bar g}\,\bar g^{ab}\partial_b\Phi\right)=0.
\end{equation}
Its first nontrivial order fixes
\begin{equation}\label{eq:psi2}
    \widetilde\phi_{(2)}=-\frac{\ell^2}{4}D^2\phi\,.
\end{equation}
The normal and mixed Einstein equations are, respectively,
\begin{align}
    K^2-K_{ab}K^{ab}-\mathcal R[\mathrm g]
    &=\frac{2}{\ell^2}+\frac{\kappa^2}{e^2}
    \left[(n\cdot\nabla\Phi)^2-(\partial\Phi)^2_{\mathrm g}\right],\nonumber\\
    D^{(\mathrm g)}_b(K^b{}_a-K\delta^b{}_a)
    &=\frac{\kappa^2}{e^2}(n\cdot\nabla\Phi)\partial_a\Phi\,.
\end{align}
Together with the tangential equations, their expansion gives
\begin{align}
    \operatorname{Tr}h_{(2)}&=0,\nonumber\\
    \operatorname{Tr}\gamma_{(2)}
    &=-\frac{\ell^2}{2}\mathcal R[\gamma]
    +\frac{\kappa^2\ell^2}{2e^2}(D\phi)^2,
    \label{eq:trace-gamma2}\\
    h_{(2)ab}
    &=-\frac{\kappa^2\ell^2}{2e^2}
    \left[D_a\phi D_b\phi-\frac12\gamma_{ab}(D\phi)^2\right],
    \label{eq:h2-full}\\
    D^b h_{(2)ab}
    &=\frac{2\kappa^2}{e^2}\widetilde\phi_{(2)}D_a\phi,
    \label{eq:constraint-divh2}\\
    D^b\!\left[\gamma_{(2)ab}
    -\gamma_{ab}\operatorname{Tr}\gamma_{(2)}\right]
    &=\frac{2\kappa^2}{e^2}\phi_{(2)}D_a\phi.
    \label{eq:constraint-divgamma2}
\end{align}
Thus the logarithmic coefficients and the trace of $\gamma_{(2)ab}$ are local functions of the sources. The remaining finite coefficients carry the state-dependent data.

\subsection{Cancellation of the on-shell divergences}
Taking the trace of the Einstein equation and substituting it into the bulk action yields
\begin{align}
    R&=-\frac6{\ell^2}+\frac{\kappa^2}{e^2}(\nabla\Phi)^2,
    \label{eq:trace-einstein-app}\\
    I_{\rm bulk}^{\varepsilon,\mathrm{on\text{-}shell}}
    &=\frac2{\kappa^2\ell^2}\int_{\mathcal M_\varepsilon}d^3x\sqrt g.
    \label{eq:on-shell-reg}
\end{align}
The expansions needed for the bulk and Gibbons--Hawking terms are
\begin{align}
    \sqrt{\mathrm g}
    &=\frac{\ell^2}{z^2}\sqrt\gamma
    \left[1+\frac{z^2}{2\ell^2}\operatorname{Tr}\gamma_{(2)}
    +O\!\left(\frac{z^4}{\ell^4}L_z^2\right)\right],\nonumber\\
    K_{ab}
    &=\frac{\ell}{z^2}\gamma_{ab}-\frac1\ell h_{(2)ab}
    +O\!\left(\frac{z^2}{\ell^3}L_z^2\right),\nonumber\\
    K&=\frac2\ell-\frac{z^2}{\ell^3}\operatorname{Tr}\gamma_{(2)}
    +O\!\left(\frac{z^4}{\ell^5}L_z^2\right).
\end{align}
Using $\sqrt g=(\ell/z)\sqrt{\mathrm g}$ and integrating over $z$ gives
\begin{align}
    I_{\rm bulk}^{\varepsilon,\mathrm{on\text{-}shell}}
    &=\frac{\ell}{\kappa^2\varepsilon^2}\int_\Sigma d^2x\sqrt\gamma
    -\frac{\log(\varepsilon/\ell)}{\kappa^2\ell}
    \int_\Sigma d^2x\sqrt\gamma\,\operatorname{Tr}\gamma_{(2)}
    +O(1),\label{eq:bulk-onshell-div}\\
    I_{\rm GH}^{\varepsilon}
    &=-\frac{2\ell}{\kappa^2\varepsilon^2}\int_\Sigma d^2x\sqrt\gamma
    +O(1).\label{eq:GH-div}
\end{align}
The trace-dependent finite terms in $\sqrt{\mathrm g}K$ cancel, so the Gibbons--Hawking term contributes no logarithmic divergence. Adding these expressions gives
\begin{equation}\label{eq:reg-div}
    I_{\rm reg}^{\rm div}
    =-\frac{\ell}{\kappa^2\varepsilon^2}\int_\Sigma d^2x\sqrt\gamma
    -\frac{\log(\varepsilon/\ell)}{\kappa^2\ell}
    \int_\Sigma d^2x\sqrt\gamma\,\operatorname{Tr}\gamma_{(2)}.
\end{equation}
Substitution of \eqref{eq:trace-gamma2} reproduces \eqref{eq:reg-div-new}, which is cancelled by the counterterms \eqref{eq:counterterm}. The logarithm uses the reference scale $\ell^{-1}$; replacing it by $\log(\varepsilon\mu_{\rm R})$ changes the finite local subtraction.

\subsection{Renormalised canonical momenta}\label{app:one-point-functions}
We vary the regulated action \eqref{eq:regulated-scalar-action} and its counterterms before using the on-shell volume expression \eqref{eq:on-shell-reg}. Define the cutoff responses by
\begin{equation}
    \delta(I_{\rm reg}+I_{\rm ct})
    =\int_{\Sigma_\varepsilon}d^2x\sqrt{\mathrm g}
    \left[-\frac12\mathcal T_\varepsilon^{ab}\delta\mathrm g_{ab}
    +\pi_\Phi^\varepsilon\delta\Phi\right].
\end{equation}
The renormalised Brown--York tensor \cite{Brown:1992br,Balasubramanian:1999re} and scalar momentum are
\begin{align}
    \mathcal T_{ab}^{\varepsilon}
    &=-\frac1{\kappa^2}
    \left(K_{ab}-K\mathrm g_{ab}+\frac1\ell\mathrm g_{ab}\right)
    +\frac\ell{e^2}\log\frac\varepsilon\ell
    \left[\partial_a\Phi\partial_b\Phi
    -\frac12\mathrm g_{ab}(\partial\Phi)^2_{\mathrm g}\right],
    \label{eq:brown-york}\\
    \pi_\Phi^\varepsilon
    &=\frac1{e^2}n\cdot\nabla\Phi
    -\frac\ell{e^2}\log\frac\varepsilon\ell\,\Box_{\mathrm g}\Phi.
    \label{eq:scalar-cutoff-momentum}
\end{align}
The curvature counterterm has only an edge variation in two dimensions, which vanishes under the boundary assumptions of \autoref{sec:Vari}. The scalar counterterm contributes to both momenta: its metric variation gives the traceless gradient term in \eqref{eq:brown-york}, while its scalar variation gives $\Box_{\mathrm g}\Phi$ after integration by parts.

To see the logarithmic cancellation explicitly, insert \eqref{eq:app-expansions}:
\begin{align}
    \mathcal T_{ab}^{\varepsilon}
    &=\frac1{\kappa^2\ell}
    \left[\gamma_{(2)ab}-\gamma_{ab}\operatorname{Tr}\gamma_{(2)}
    +h_{(2)ab}\right]\nonumber\\
    &\quad+\log\frac\varepsilon\ell
    \left\{\frac{2h_{(2)ab}}{\kappa^2\ell}
    +\frac\ell{e^2}\left[D_a\phi D_b\phi
    -\frac12\gamma_{ab}(D\phi)^2\right]\right\}+o(1),
    \label{eq:stress-log-cancellation}\\
    \frac{\ell^2}{\varepsilon^2}\pi_\Phi^\varepsilon
    &=-\frac2{e^2\ell}\left(\phi_{(2)}+\widetilde\phi_{(2)}\right)
    -\log\frac\varepsilon\ell
    \left[\frac{4\widetilde\phi_{(2)}}{e^2\ell}
    +\frac\ell{e^2}D^2\phi\right]+o(1).
    \label{eq:scalar-log-cancellation}
\end{align}
The logarithmic coefficients vanish by \eqref{eq:h2-full} and \eqref{eq:psi2}. In boundary dimension two, the lower-index stress tensor needs no further Weyl rescaling, whereas the scalar response carries the measure factor $\ell^2/\varepsilon^2$. Thus
\begin{equation}
    T_{ab}=\lim_{\varepsilon\to0}\mathcal T_{ab}^{\varepsilon},
    \qquad
    \psi=\lim_{\varepsilon\to0}\frac{\ell^2}{\varepsilon^2}\pi_\Phi^\varepsilon.
\end{equation}

\paragraph{Finite responses and Ward identities.}\label{app:response-variables}
The finite coefficients are related to the responses by
\begin{align}
    T_{ab}
    &=\frac1{\kappa^2\ell}
    \left[\gamma_{(2)ab}-\gamma_{ab}\operatorname{Tr}\gamma_{(2)}
    +h_{(2)ab}\right],\label{eq:response-dictionary}\\
    \psi
    &=-\frac2{e^2\ell}\left(\phi_{(2)}+\widetilde\phi_{(2)}\right).
    \label{eq:one-point}
\end{align}
Solving these relations for $\gamma_{(2)ab}$ and $\phi_{(2)}$ gives the parametrisation in \eqref{eq:asymptotic-metric-new} and \eqref{eq:asymptotic-phi-new}. Taking the trace of \eqref{eq:response-dictionary} and using \eqref{eq:trace-gamma2} yields \eqref{eq:Weyl-anomaly-new}. Its divergence, together with \eqref{eq:constraint-divh2} and \eqref{eq:constraint-divgamma2}, gives \eqref{eq:div-constraint}. These checks identify the independent radial data with the renormalised one-point functions in the main text.

\section{Charge variation and spatial endpoints}\label{app:charge-variation}
The charged-brane family has scalar source $\phi=-e^2qx$ and local response $\psi=0$. Varying $q$ changes the source by $\delta\phi=-e^2x\,\delta q$, including its values at spatial infinity. We evaluate this variation by temporarily restricting the solution to $-W<x<W$ and $r\leq r_c$, retaining the induced data at the spatial faces. The charge derivative is taken at fixed $\beta$, $W$, $\ell$ and $e$.

For the metric in \autoref{sec:charged-brane}, $\sqrt g\,g^{xx}=\ell/r$. The scalar kinetic term therefore contributes
\begin{align}\label{eq:scalar-end-flux}
\mathcal B_\Phi(r_c)
&=\frac1{e^2}\int_0^\beta d\tau\int_{r_h}^{r_c}dr\,
\left[\sqrt g\,g^{xx}\partial_x\Phi\,\delta\Phi\right]_{-W}^{W}\nonumber\\
&=2\beta We^2\ell q\log\frac{r_c}{r_h}\,\delta q.
\end{align}
The radial scalar response vanishes because $\partial_r\Phi=0$. At the asymptotic boundary, the logarithmic scalar counterterm gives
\begin{equation}
\mathcal B_{\rm ct}(r_c)
=-2\beta We^2\ell q\log\frac{r_c}{\ell}\,\delta q+o(1).
\end{equation}
Combining these terms and removing the radial cutoff yields
\begin{equation}\label{eq:charge-endpoint-response}
\left.\delta I_{\rm ren}\right|_{\beta,W}
=\lim_{r_c\to\infty}\bigl(\mathcal B_\Phi+\mathcal B_{\rm ct}\bigr)
=-2\beta We^2\ell q\log\frac{r_h}{\ell}\,\delta q.
\end{equation}
Direct differentiation of the saddle action \eqref{eq:canonical-saddle-action}, with $r_h=r_h(\beta,q)$, gives the same result.

In Maxwell variables, the extensive charge and renormalised chemical potential are
\begin{equation}
Q=2Wq,\qquad
\mu=-e^2\ell q\log\frac{r_h}{\ell}.
\end{equation}
Indeed, subtracting the logarithm in \eqref{eq:brane-gaugefield} at scale $\ell$ gives $A_\tau^{\rm ren}=i e^2\ell q\log(r_h/\ell)$, with $\mu=iA_\tau^{\rm ren}$ in our Euclidean continuation. For variations along this homogeneous family, the endpoint term per unit Euclidean time can be represented by the distribution
\begin{equation}
\psi_{\rm end}^{(W)}(x)=-\frac{\mu}{e^2}
\bigl[\delta(x-W)-\delta(x+W)\bigr].
\end{equation}
Its pairing with $\delta\phi=-e^2x\,\delta q$ gives $\delta I_{\rm ren}/\beta=2W\mu\,\delta q$. For any fixed smooth test variation of compact spatial support, both endpoint values vanish for sufficiently large $W$. Thus $\psi_{\rm end}^{(W)}$ tends to zero as a distribution on the open line, while the slope variation gives a finite response per unit length. The finite counterterm of \autoref{subsec:holo-ren-onepoint} shifts the action by $2\alpha\beta We^4q^2$ and the conjugate potential by $\Delta\mu=2\alpha e^4q$.

\section{The quadratic kernel and the annular source pairing}
\label{app:quadratic-prescription}
This appendix fixes the finite quadratic subtraction and the treatment of the logarithmic source used in \autoref{sec:quadratic-revision}. We first determine the renormalised kernel for smooth sources and then evaluate the unbounded profile with explicit radial and spatial regulators.

\paragraph{The renormalised kernel.}
In physical Cartesian coordinates, use the Fourier convention
\begin{equation}
\phi(x)=\int\frac{d^2k}{(2\pi)^2}e^{ik\cdot x}\phi(k),
\qquad k=|\boldsymbol k|.
\end{equation}
The regular Poincar\'e solution is $\Phi(z,k)=kzK_1(kz)\phi(k)$, where $K_1$ is the modified Bessel function. Its boundary expansion is
\begin{equation}
\Phi(z,k)=\left\{1+\frac{k^2z^2}{2}
\left[\log\frac{kz}{2}+\gamma_{\rm E}-\frac12\right]
+O(z^4\log z)\right\}\phi(k).
\end{equation}
For smooth sources, the quadratic action follows from the scalar kinetic term on the neutral background. The metric backreaction starts at order $\phi^2$, and its linear contribution to the gravitational action vanishes by stationarity at fixed boundary metric. With the outward normal pointing towards decreasing $z$, the radial scalar action and the counterterm \eqref{eq:counterterm} therefore give
\begin{align}
I_{\Phi,\rm rad}^{(2)}(\varepsilon)
&=-\frac{\ell}{2e^2}\int\frac{d^2k}{(2\pi)^2}
\phi(k)\phi(-k)k^2
\left[\log\frac{k\varepsilon}{2}+\gamma_{\rm E}\right]+o(1),\\
I_{\rm ct,\Phi}^{(2)}(\varepsilon)
&=\frac{\ell}{2e^2}\log\frac{\varepsilon}{\ell}
\int\frac{d^2k}{(2\pi)^2}\phi(k)\phi(-k)k^2+o(1).
\end{align}
Adding $I_{\rm fin}=\alpha\int(\partial\phi)^2$ and using $\mathcal G=-I_{\rm ren}$ yields
\begin{align}
\mathcal G^{(2)}
&=\frac12\int\frac{d^2k}{(2\pi)^2}
\phi(k)\phi(-k)\widetilde K_\alpha(k),\nonumber\\
\widetilde K_\alpha(k)
&=\frac{\ell}{2e^2}k^2\log(k^2\ell^2)
+\left[\frac{\ell}{e^2}(\gamma_{\rm E}-\log2)-2\alpha\right]k^2.
\label{eq:quadratic-fourier-kernel}
\end{align}
The inverse transform at separated points is $C_{\mathcal O_\phi}/|x|^4$. The polynomial term specifies its local contact extension. In particular, $I_{\rm fin}=0$ selects the displayed constant $\gamma_{\rm E}-\log2$, with reference length $\ell$.

\paragraph{The logarithmic source.}
To define its plane pairing, use dimensionless Poincar\'e coordinates,
\begin{equation}
ds^2=\frac{\ell^2}{u^2}(du^2+d\rho^2+\rho^2d\theta^2),
\qquad
\Phi(u,\rho)=-\frac b2\log(u^2+\rho^2).
\end{equation}
This profile solves the scalar equation and tends to $\phi=-b\log\rho$. We regulate the domain by $u\geq\epsilon$ and $\rho_-\leq\rho\leq\rho_+$, retain the values induced by this solution on both radial sides of the annulus, and take $\epsilon\to0$ at fixed $\rho_\pm$. Here $\epsilon=z/\ell$ is dimensionless. The scalar kinetic integral is
\begin{align}\label{eq:annular-scalar-bulk}
I_{\Phi,\rm bulk}^{(2)}
&=\frac{\pi\ell b^2}{e^2}
\int_{\rho_-}^{\rho_+}\rho\,d\rho
\int_\epsilon^\infty\frac{du}{u(u^2+\rho^2)}\nonumber\\
&=\frac{\pi\ell b^2}{e^2}
\int_{\rho_-}^{\rho_+}\frac{d\rho}{\rho}
\log\frac{\sqrt{\epsilon^2+\rho^2}}{\epsilon}.
\end{align}
The scalar counterterm is
\begin{equation}
I_{\rm ct,\Phi}^{(2)}
=\frac{\pi\ell b^2}{e^2}\log\epsilon
\int_{\rho_-}^{\rho_+}d\rho\,
\frac{\rho^3}{(\epsilon^2+\rho^2)^2}.
\end{equation}
Their sum has the finite limit
\begin{equation}\label{eq:annular-scalar-finite}
I_{\Phi,\rm ren}^{(2)}
=\frac{\pi\ell b^2}{2e^2}
\left[(\log\rho_+)^2-(\log\rho_-)^2\right]
+2\pi\alpha b^2\log\frac{\rho_+}{\rho_-}.
\end{equation}
This evaluation includes the spatial end terms. Explicitly, integration by parts separates the kinetic action into the $u=\epsilon$ face and the two faces $\rho=\rho_\pm$. Writing $L_\pm=\log\rho_\pm$, their contributions are
\begin{align}
I_{\Phi,\rm rad}^{(2)}
&=-\frac{\pi\ell b^2}{2e^2}(L_+^2-L_-^2)+o(1),\\
I_{\Phi,\rm ends}^{(2)}
&=\frac{\pi\ell b^2}{e^2}
\left[L_+^2-L_-^2-\log\epsilon\,(L_+-L_-)\right]+o(1).
\end{align}
The face at $u\to\infty$ contributes zero. The logarithm in the end contribution cancels against the counterterm, reproducing \eqref{eq:annular-scalar-finite}. For $\rho_\pm=e^{\pm\mathcal L}$ and $\alpha=0$, the finite plane pairing consequently vanishes. At nonzero $\alpha$, the shift is $4\pi\alpha b^2\mathcal L=2\alpha\beta We^4q^2$, as in the cylinder calculation. The inner circle regulates the puncture at $\rho=0$; its contribution has been retained before the limit $\mathcal L\to\infty$. Equations \eqref{eq:quadratic-fourier-kernel} and \eqref{eq:annular-scalar-finite} specify, respectively, the local subtraction and the annular extension used in the quadratic comparison.

\section{Shadow normalisations and the quartic source vertices}
\label{app:quartic-details}
We give the normalisation and source-integral identities used in \autoref{sec:quartic-revision}, including the coupling conversion for the bulk exchange coefficients, the block convention and the evaluation of the logarithmic pair vertices.

\paragraph{Exchange coefficients and gravitational coupling.}
For boundary dimension $d=2$ and external dimension $\Delta_{\mathcal O_\phi}=2$, the graviton-exchange coefficients for unit-normalised external operators take the form \cite{Costa:2014kfa}
\begin{align}\label{eq:exchange-coefficients-coupling}
    \widetilde b_2(\nu)
    &=\frac{\kappa^2}{2\pi\ell}\frac1{\nu^2+1}
    =\frac1c\frac6{\nu^2+1}\,,\nonumber\\
    \widetilde b_0(\nu)
    &=-\frac{\kappa^2}{64\pi\ell}
    \frac{(\nu^2+1)^2}{\nu^2+4}
    =-\frac1c\frac3{16}\frac{(\nu^2+1)^2}{\nu^2+4}\,.
\end{align}
Here $\kappa^2=8\pi G_3$ and $c=12\pi\ell/\kappa^2$. The numerators follow from the two scalar--graviton vertices, while the denominators are those of the spin-two and trace terms of the split graviton propagator. Factoring out $1/c$ as in \eqref{eq:quartic-kernel-definition} gives $b_j=c\widetilde b_j$ and hence \eqref{eq:bulk-spectral-coefficients}.

\paragraph{Complex beta identity and the raw OPE coefficients.}
For a pair $\mathbf a=(a|\bar a)$ with $a-\bar a\in\mathbb Z$, define
\begin{equation}
    \zeta^{\mathbf a}:=\zeta^a\bar\zeta^{\bar a}\,,\qquad
    \Gamma^{\mathbb C}(a)=i^{a-\bar a}
    \frac{\Gamma(a)}{\Gamma(1-\bar a)}\,.
\end{equation}
In expressions involving paired exponents, arithmetic is componentwise: $y^{a-1}$ denotes $y^{a-1}\bar y^{\bar a-1}$. With the area measure used in \eqref{eq:raw-shadow}, the complex beta identity reads \cite{Neretin_2023}
\begin{equation}\label{eq:complex-beta-revision}
    \operatorname{AC}\int_{\mathbb C}d^2y\,
    y^{a-1}(1-y)^{b-1}
    =\pi\frac{\Gamma^{\mathbb C}(a)\Gamma^{\mathbb C}(b)}
    {\Gamma^{\mathbb C}(a+b)}\,.
\end{equation}
The integral is initially evaluated in its convergence domain and then continued.

Choose the conformal frame $(\zeta_1,\zeta_2,\zeta_3,\zeta_4)=(\zeta,0,1,\infty)$ and remove the external prefactor. The physical OPE branch of the raw wave \eqref{eq:raw-shadow} is $\zeta^h\bar \zeta^{\bar h}$ with coefficient
\begin{align}\label{eq:raw-ope-coefficient}
    S(h,\bar h)
    &=\operatorname{AC}\int d^2y\,y^{-2h}(1-y)^{h-1}
    =\pi\frac{\Gamma^{\mathbb C}(1-2h)\Gamma^{\mathbb C}(h)}
    {\Gamma^{\mathbb C}(1-h)}\nonumber\\
    &=\pi\frac{\Gamma(1-2h)\Gamma(h)\Gamma(\bar h)}
    {\Gamma(2\bar h)\Gamma(1-h)\Gamma(1-\bar h)}\,.
\end{align}
The integral in the first line uses paired powers. The second OPE region gives the shadow coefficient $S(1-h,1-\bar h)$. For the even helicities considered here, the phases from reversing coordinate differences cancel. Writing $a=(1+i\nu)/2$ gives
\begin{equation}\label{eq:raw-spin-coefficients}
    S_0=\pi\frac{\Gamma(1-2a)\Gamma(a)^2}
    {\Gamma(2a)\Gamma(1-a)^2}\,,\qquad
    S_2=\pi\frac{\Gamma(3-2a)\Gamma(a-1)^2}
    {\Gamma(2a+2)\Gamma(-a)^2}\,.
\end{equation}
The coefficient for helicity $-2$ is the same as for $+2$.

\paragraph{Conversion to normalised partial waves.}
The reduced four-point function uses the cross-ratios $\zeta=\zeta_{12}\zeta_{34}/(\zeta_{13}\zeta_{24})$ and its antiholomorphic counterpart $\bar\zeta$. Define $k_a(\zeta)=\zeta^a{}_2F_1(a,a;2a;\zeta)$. We use the two-dimensional global blocks \cite{Costa:2012cb}
\begin{align}\label{eq:quartic-global-blocks}
    G_{\Delta,0}&=k_{\Delta/2}(\zeta)k_{\Delta/2}(\bar \zeta)\,,\nonumber\\
    G_{\Delta,2}&=\frac14\left[
    k_{(\Delta+2)/2}(\zeta)k_{(\Delta-2)/2}(\bar \zeta)
    +k_{(\Delta-2)/2}(\zeta)k_{(\Delta+2)/2}(\bar \zeta)\right].
\end{align}
The factor $1/4$ specifies our spin-two OPE normalisation. The corresponding block-plus-shadow wave is
\begin{equation}\label{eq:quartic-block-shadow}
  F_{\nu,j}
= \mathcal{N}_{\nu,j} G_{1+i\nu,j}
+ \mathcal{N}_{-\nu,j} G_{1-i\nu,j},
\qquad
\mathcal{N}_{\nu,j} = \frac{i\nu}{2\pi K_{1+i\nu,j}}.
\end{equation}
where, for four identical external operators of dimension two, the normalisation coefficient is
\begin{equation}\label{eq:partial-wave-K}
    K_{\Delta,j}=
    \frac{\Gamma(\Delta+j)\Gamma(\Delta)(\Delta-1)_j}
    {4^{j-1}\Gamma^4((\Delta+j)/2)
    \Gamma^2((4-\Delta+j)/2)\Gamma^2((2+\Delta+j)/2)}\,.
\end{equation}
Here $(a)_j=\Gamma(a+j)/\Gamma(a)$ is the Pochhammer symbol. The full wave $\mathcal F_{\nu,j}=|\zeta_{12}|^{-4}|\zeta_{34}|^{-4}F_{\nu,j}$ includes the external factor. Matching the coefficients of the physical and shadow OPE branches gives
\begin{equation}
    B_0=\frac{S_0}{\mathcal{N}_{\nu,0}}\,,\qquad
    B_2=\frac{4S_2}{\mathcal{N}_{\nu,2}}\,.
\end{equation}
The factor four in $B_2$ compensates the block normalisation above, whereas the sum of two raw helicity waves appears explicitly in \eqref{eq:shadow-wave-conversion}.

Gamma recurrence and reflection reduce these ratios to
\begin{align}
    B_0&=\frac{8\pi^2\Gamma(i\nu)\Gamma(-i\nu)}
    {[\Gamma(a)\Gamma(1-a)]^2[\Gamma(1+a)\Gamma(2-a)]^2}\,,\nonumber\\
    B_2&=\frac{2\pi^2\nu^2(1+\nu^2)\Gamma(i\nu)\Gamma(-i\nu)}
    {[\Gamma(1+a)\Gamma(2-a)]^2[\Gamma(2+a)\Gamma(3-a)]^2}\,.
\end{align}
Using
\begin{align}
    \Gamma(i\nu)\Gamma(-i\nu)&=\frac{\pi}{\nu\sinh\pi\nu}\,,&
    \Gamma(a)\Gamma(1-a)&=\frac{\pi}{\cosh(\pi\nu/2)}\,,\nonumber\\
    \Gamma(1+a)\Gamma(2-a)&=\frac{\pi(1+\nu^2)}{4\cosh(\pi\nu/2)}\,,&
    \Gamma(2+a)\Gamma(3-a)&=\frac{\pi(1+\nu^2)(9+\nu^2)}{16\cosh(\pi\nu/2)}\,,
\end{align}
we obtain \eqref{eq:shadow-normalisations-revision}. Singular individual factors at $\nu=0$ are understood by taking the limit of the complete spectral integrand.

The stress-tensor normalisation can also be checked directly. With $\langle T(\zeta)T(0)\rangle=c/(2\zeta^4)$ and external weights $(1,1)$, the Ward identity gives a coefficient $2/c$ for each chiral stress-tensor block. Their contribution to the reduced correlator is therefore
\begin{equation}\label{eq:stress-block-check}
    \frac2c\left[k_2(\zeta)+k_2(\bar\zeta)\right]
    =\frac8c\,G_{2,2}(\zeta,\bar\zeta)\,,
    \qquad k_0(\zeta)=1\,.
\end{equation}
Equation \eqref{eq:partial-wave-K} gives $K_{2,2}=3/4$, so the spin-two pole coefficient is $K_{2,2}(8/c)=6/c$, in agreement with \eqref{eq:exchange-coefficients-coupling}.

\paragraph{Constant leg and logarithmic source vertex.}
Integrating the first leg of \eqref{eq:shadow-three-point-revision} with a constant source and setting $u=(\zeta_1-y)/(\zeta_2-y)$ leaves the beta integral
\begin{equation}\label{eq:quartic-constant-leg}
    \operatorname{AC}\int d^2\zeta_1\,\mathcal T_{h,\bar h}(12;y)
    \ \propto\ 
    \frac{\Gamma^{\mathbb C}(1-h)\Gamma^{\mathbb C}(h-1)}
    {\Gamma^{\mathbb C}(0|0)}=0\,.
\end{equation}
This holds at generic principal-series weights, with exceptional values defined by continuation. It removes the constant shifts of the logarithms under $\zeta_i=y u_i$.

The source integrals are of Dotsenko--Fateev type \cite{Dotsenko:1984ad,Neretin_2023}. For the logarithmic source at $y=1$, set $t_i=(1-\zeta_i)^{-1}$ and write $L(t)=\log|t-1|-\log|t|$. The Jacobians cancel the denominators in \eqref{eq:shadow-three-point-revision}. With $t_{12}=t_1-t_2$, the vertex becomes
\begin{equation}
    v_j=\operatorname{AC}\int d^2t_1\,d^2t_2\,
    L(t_1)L(t_2)\,t_{12}^{h-2}\bar t_{12}^{\bar h-2}.
\end{equation}
Let $H(t)=t^h\bar t^{\bar h}$, $D=h\bar h(h-1)(\bar h-1)$ and $\Delta_i=4\partial_{t_i}\partial_{\bar t_i}$. The continued kernel is $\Delta_1\Delta_2 H(t_{12})/(16D)$. Transferring the Laplacians to the sources gives $\Delta L=2\pi(\delta_1-\delta_0)$, where $\delta_a$ is the plane delta function at $a$. For $0<\operatorname{Re}(h+\bar h)<1$, $H(0)=0$ and $H(\pm1)=1$ for even spin. Thus
\begin{equation}
    v_j=\frac{(2\pi)^2}{16D}\bigl[2H(0)-H(1)-H(-1)\bigr]
    =-\frac{\pi^2}{2D}.
\end{equation}
The integrations by parts can be taken with a smooth cutoff at radius $R$. Since $L=O(R^{-1})$ and $\nabla L=O(R^{-2})$, the cutoff terms vanish in this strip as $R\to\infty$. Continuing to $h+\bar h=1+i\nu$ gives \eqref{eq:source-vertices-result}. The shadow weights $(1-h,1-\bar h)$ give the same $D$ and hence the same vertex.

\phantomsection
\addcontentsline{toc}{section}{References}
\bibliographystyle{fullsort}
\bibliography{references}
\end{document}